\documentclass[12pt]{article}
\usepackage{a4wide}
\usepackage{amssymb}
\usepackage{amsmath}
\usepackage{graphicx}
\usepackage{slashed}
\usepackage{verbatim}
\usepackage{hyperref}

\def\Xint#1{\mathchoice
   {\XXint\displaystyle\textstyle{#1}}%
   {\XXint\textstyle\scriptstyle{#1}}%
   {\XXint\scriptstyle\scriptscriptstyle{#1}}%
   {\XXint\scriptscriptstyle\scriptscriptstyle{#1}}%
   \!\int}
\def\XXint#1#2#3{{\setbox0=\hbox{$#1{#2#3}{\int}$}
     \vcenter{\hbox{$#2#3$}}\kern-.52\wd0}}

\def\dashint{\Xint-}

\newcommand{\be}{\begin{equation}}\newcommand{\ee}{\end{equation}}
\newcommand{\bea}{\begin{eqnarray}} \newcommand{\eea}{\end{eqnarray}}

\def\aaa{\sigma }

\def\makeatletter{\catcode`\@=11}% 11:letter
\makeatletter
\def\mathbox#1{\hbox{$\m@th#1$}}%
\def\math@ccstyles#1#2#3#4#5#6#7{{\leavevmode
      \setbox0\mathbox{#6#7}%
      \setbox2\mathbox{#4#5}%
      \dimen@ #3%
      \baselineskip\z@\lineskiplimit#1\lineskip\z@
      \vbox{\ialign{##\crcr
             \hfil \kern #2\box2 \hfil\crcr
             \noalign{\kern\dimen@}%
             \hfil\box0\hfil\crcr}}}}
\def\mathaccstyles{\math@ccstyles\maxdimen}
\def\maththroughstyles{\math@ccstyles{-\maxdimen}}
\def\unity%
 {\maththroughstyles{.45\ht0}\z@\displaystyle {\mathchar"006C}\displaystyle 1}

\makeatletter \@addtoreset{equation}{section} \makeatother

\begin{document}

\setcounter{table}{0}

\begin{flushright}\footnotesize

\texttt{ICCUB-16-017}
\vspace{0.6cm}
\end{flushright}

\mbox{}
\vspace{0truecm}
\linespread{1.1}

%%%%%%%%%%%%%%%%%
\centerline{\LARGE \bf Large $N$ Correlation Functions  }

\vspace{.5cm}

 \centerline{\LARGE \bf in Superconformal Field Theories}

\vspace{1.5truecm}

\centerline{
    {\large \bf Diego Rodriguez-Gomez${}^{a}$} \footnote{d.rodriguez.gomez@uniovi.es}
    {\bf and}
    {\large \bf Jorge G. Russo ${}^{b,c}$} \footnote{jorge.russo@icrea.cat}}

\vspace{1cm}
\centerline{{\it ${}^a$ Department of Physics, Universidad de Oviedo}} \centerline{{\it Avda.~Calvo Sotelo 18, 33007  Oviedo, Spain}}
\medskip
\centerline{{\it ${}^b$ Instituci\'o Catalana de Recerca i Estudis Avan\c{c}ats (ICREA)}} \centerline{{\it Pg.Lluis Compayns, 23, 08010 Barcelona, Spain}}
\medskip
\centerline{{\it ${}^b$ Departament de F\' \i sica Cu\' antica i Astrof\'\i sica and Institut de Ci\`encies del Cosmos}} \centerline{{\it Universitat de Barcelona}}\centerline{{\it Mart\'i Franqu\`es, 1, 08028
Barcelona, Spain }}
\vspace{1cm}

\centerline{\bf ABSTRACT}
\medskip

We compute correlation functions of chiral primary operators in $\mathcal{N}=2$ superconformal theories at large $N$ using a construction based on supersymmetric localization recently developed by Gerchkovitz \textit{et al.} We focus on $\mathcal{N}=4$ SYM as well as on superconformal QCD. In the case of $\mathcal{N}=4$ we recover the free field theory results as expected due to non-renormalization theorems. In the case of superconformal QCD we study the planar expansion in the large $N$ limit. The final correlators admit a simple generalization to a finite $N$ formula which exactly matches the various small $N$ results in the literature.

\noindent

\newpage

\section{Introduction}

Supersymmetric gauge theories with $\mathcal{N}=2$ superconformal invariance are of great interest, as they provide a tractable, yet very rich, class of theories. In this respect, the recently developed new tools, including in particular supersymmetric localization \cite{Pestun:2007rz}, allow to perform exact computations of a plethora of observables, including for instance the partition function, Wilson and 't Hooft loops \cite{Pestun:2007rz,Gomis:2011pf,Rey:2010ry,Passerini:2011fe,Bourgine:2011ie,Russo:2012ay}, domain walls   \cite{Drukker:2010jp,Hosomichi:2010vh} or cusp anomalous dimensions  \cite{Fiol:2015spa,Fiol:2015mrp}. 

A particularly interesting sector of $\mathcal{N}=2$ superconformal theories (SCFT's) is that originated from primary operators annihilated by all supercharges of one chirality, hence known as chiral primaries. In superspace language, the scaling dimension $\Delta$ of  chiral primaries are bottom components of $\mathcal{N}=2$ chiral superfields. The case of $\Delta=2$ is particularly interesting, as the integrated top component of the multiplet defines an exactly marginal operator. Hence, chiral superfields with $\Delta=2$ parametrize the conformal manifold associated to the SCFT. Moreover, the 2-point function of such top components defines the Zamolodchikov metric on the conformal manifold. Such metric depends on the exactly marginal couplings, which act then as coordinates. Because of supersymmetry, it turns out that such metric can be read as well from the correlators of the chiral primary operators (CPO's). 

In general, supersymmetry guarantees that the OPE of CPO's is non-singular. Thus, they are endowed with a ring structure, the so-called chiral ring. While there is no strict proof, it is believed that this ring is spanned by a finite set of generators. As one moves on the space of exactly marginal couplings, CPO's generically mix. Thus, CPO's can be thought as sections of a bundle endowed with a connection encoding such mixings. Similarly to the 2d case \cite{Witten:1989ig,Dijkgraaf:1990dj,Dijkgraaf:1990qw,Cecotti:1991me}, it turns out that the integrability conditions for this connection define a set of $tt^*$ equations which, together with the Witten-Dijkgraaf-Verlinde-Verlinde equations, encode the (holomorphic) coupling dependence of the correlation functions \cite{Papadodimas:2009eu,Baggio:2014ioa,Baggio:2015vxa}.

Very recently, it has been shown that the Zamolodchikov metric on the conformal manifold can be exactly computed by means of supersymmetric localization. In particular, it turns out that the $S^4$ partition function can be identified with the Kahler potential for the Zamolodchikov metric  \cite{Gerchkovitz:2014gta,Gomis:2014woa}. Then, the derivatives of the partition function with respect to the marginal couplings allow to compute correlation functions for exactly marginal operators.

In \cite{Gerchkovitz:2016gxx} this result was extended and a method to exactly compute correlation functions of arbitrary CPO's was developed. The starting point is a modified version of the theory on the $S^4$ whereby it is deformed by couplings to all the generators of the chiral ring --that is, for each CPO generating the chiral ring, one adds to the theory the integrated top component of the superfield whose bottom component is that CPO. By taking derivatives with respect to these couplings and upon setting at the end the deformation couplings to zero, one can compute correlation functions of the associated CPO's. The important caveat noticed in \cite{Gerchkovitz:2016gxx} is that when going from the $S^4$ into $\mathbb{R}^4$, due to the conformal anomaly further mixings are introduced \cite{Gomis:2015yaa}. For instance, while on $\mathbb{R}^4$ the correlator of two CPO's with different scaling dimensions automatically vanishes due to Ward identities, on the $S^4$ this is not anymore the case; as the conformal anomaly allows for a mixing among operators of different dimension (basically because the Ricci scalar, being dimensionful, allows to mix operators with dimensions differing by 2). Thus, in order to disentangle this mixing, one should perform, on the $S^4$ result, a Gram-Schmidt orthogonalization.

In this paper we will apply the prescription given in \cite{Gerchkovitz:2016gxx} to compute correlators of CPO's to $\mathcal{N}=4$ SYM and superconformal QCD. Since these theories admit a lagrangian description, we can identify the CPO's of interest with gauge-invariant operators involving only the scalar field in the vector multiplet $\phi$. Moreover, we will be interested in the large $N$ limit of these theories, where we can use the saddle point approximation to compute the deformed partition function and its derivatives. The large $N$ limit introduces further simplifications, as multi-traces, having an extra $1/N$ suppression, decouple. Hence, in this limit, the CPO's of interest on $\mathbb{R}^4$ are simply ${\rm Tr}\phi^n$ for $n$ up to $N$. In addition, the large $N$ limit has the effect to  suppress instanton contributions as well. While \textit{a priori} it might be that the integration over instanton moduli could overcome the exponential suppression of the instanton action, it is widely believed that in practise the moduli space factor is at most power-like in $N$, and thus cannot compensate the exponential damping (see \cite{Passerini:2011fe} for an explicit test). Thus, in the large $N$ limit we can drop the instanton contribution which would, otherwise, be present for the superconformal QCD theory. While in large $N$ one would naively expect that $U(N)$ and $SU(N)$ are indistinguishable, in this context there is one important caveat: as opposed to the $SU(N)$ case, in the $U(N)$ theory the operator ${\rm Tr}\phi$ is present. This is relevant as this operator will mix on the $S^4$, due to the conformal anomaly, with all operators of the form ${\rm Tr}\phi^{2n+1}$. In most of the discussion we will assume that the gauge group is  $U(N)$. The case of $SU(N)$ gauge group will be treated  at the end of  in section \ref{N=4}.

This paper is organized as follows: in section \ref{review} we review some salient aspects of the construction in \cite{Gerchkovitz:2016gxx}. In section \ref{N=4} we consider  the case of the $\mathcal{N}=4$ theory at large $N$. Using saddle point techniques, we compute the relevant derivatives of the partition function and perform the orthogonalization to obtain the desired correlators on $\mathbb{R}^4$. In section \ref{sect:SQCD} we turn to the large $N$ limit of superconformal QCD. In this case, due to the non-trivial one-loop determinant in the $S^4$ partition function, we compute the two-point functions in the  weak coupling expansion in the 't Hooft coupling (planar expansion) and briefly comment on the strong coupling regime. Finally, in section \ref{conclusions} we offer some conclusions and highlight some open problems. For completeness, we discuss the decoupling of multi-trace operators in appendix \ref{DecouplingOfMultitrace}; and compile several useful technical results in appendix \ref{details}. In appendix \ref{HigherDerivatives} we compute higher derivatives of the free energy, which could be of interest for higher-point function computations.

\section{Exact correlators for chiral primary operators in $\mathbb{R}^4$ from the $S^4$ partition function}\label{review}

Local operators in SCFT's are organized into highest weight representations of the conformal algebra. These are labelled by their highest weight state $O$ under dilatations, known as the superconformal primary. In turn, $O$ can be defined as $[S^a_{\alpha},O]=[\overline{S}^a_{\dot{\alpha}},O]=0$, where $S^a_{\alpha}$ and $\overline{S}^a_{\dot{\alpha}}$ are the conformal supersymmetry generators. Among all superconformal primaries, a particularly interesting class is that of chiral primary operators, defined as $[\overline{Q}^a_{\dot{\alpha}},O]=0$, where $Q^a_{\alpha}$ and $\overline{Q}^a_{\dot{\alpha}}$ are the Poincare supercharges. Anti-chiral primary operators are defined analogously as $[Q^a_{\alpha},\overline{O}]=0$.

There are strong arguments suggesting that chiral primaries are always Lorentz scalars satisfying the BPS bound $\Delta_{O}=\frac{R_{O}}{2}$ (or $\Delta_{\overline{O}}=-\frac{R_{\overline{O}}}{2}$ for antichirals) --see \cite{Buican:2014qla,Gerchkovitz:2016gxx} for comments on this issue. Because of this BPS bound, the OPE of  CPO is non-singular, and the structure functions in the OPE become simply constants. This implies that these operators are endowed with a ring structure which is expected to be freely generated and with dimension equal to the dimension of the Coulomb branch of the theory when a Lagrangian description is available. Thus, one may choose a basis where the 2-point functions of CPO's on $\mathbb{R}^4$ are of the form

\begin{equation}
\label{R4}
\langle O_n(x)\overline{O}_{\overline{m}}(0)\rangle_{\mathbb{R}^4}=\frac{G_{n\overline{m}}}{|x|^{2\Delta_n}}\delta_{\Delta_n\Delta_{\overline{m}}}\ .
\end{equation}
Then, the metric $G_{n\overline{m}}$ encodes all the essential data about the chiral ring, and it is the main object of interest in this paper. 

Very recently, \cite{Gerchkovitz:2016gxx} suggested that correlators of CPO's --and hence the metric $G_{n\overline{m}}$ in eq.\eqref{R4}-- can be extracted from the $S^4$ partition function of an auxiliary theory, the latter computable in principle exactly due to supersymmetric localization. More explicitly, in superfield language, the superfield whose bottom component is a CPO $O$ with conformal dimension $\Delta$ --denoted in an abuse of notation by $O$ as well-- satisfies that $\overline{D}^a_{\dot{\alpha}}O$. Following \cite{Gerchkovitz:2016gxx}, one can consider a deformation of the $S^4$ theory preserving $osp(2|4)$, namely the supergroup of the general massive theory. In superspace, the deformed SCFT on the $S^4$ is constructing by deforming with

\begin{equation}
-\frac{1}{32\pi^2}  \int d^4x\int d^4\theta \mathcal{E}\,\tau_{O}\,O\ ,
\end{equation}
where $\mathcal{E}$ is the chiral density. This way one can construct a deformed $S^4$ partition function $\mathcal{Z}(\{\tau_n,\overline{\tau}_{\overline{n}}\})$ depending on all the deformation couplings $\tau_n$ --note that some of them may have $\Delta=2$ and  thus correspond to exactly marginal operators parametrizing the conformal manifold. As shown in \cite{Gerchkovitz:2016gxx}, it turns out then that 

\begin{equation}
\frac{1}{\mathcal{Z}(\tau_n,\overline{\tau}_{\overline{n}})} \partial_{\tau_n}\partial_{\overline{\tau}_{\overline{m}}}\mathcal{Z}(\tau_n,\overline{\tau}_{\overline{n}})= \Big(\frac{1}{32\pi^2}\Big)^2 \int d^4x\sqrt{g(x)}\int d^4y \sqrt{g(y)} \langle \mathcal{C}_n(x)\overline{C}_{\overline{m}}(y)\rangle_{S^4}\ ,
\end{equation}
where $\mathcal{C}_n$ is a quantity constructed out of the CPO, of the top component of the superfield $C$ and of the middle component $B$ (see \cite{Gerchkovitz:2016gxx} for details). Using for instance a Ward identity, it then follows that the integrated correlator of the $\mathcal{C}_n$'s equals the correlation function of the CPO's $\mathcal{O}$ evaluated at the north and south poles of the sphere

\begin{equation}
\Big(\frac{1}{32\pi^2}\Big)^2 \int d^4x\sqrt{g(x)}\int d^4y \sqrt{g(y)} \langle \mathcal{C}_n(x)\overline{C}_{\overline{m}}(y)\rangle_{S^4}=\langle O_n(N)\overline{O}_{\overline{m}}(S)\rangle_{S^4}\ .
\end{equation}
Thus the correlator on the $S^4$ is

\begin{equation}
\langle O_n(N)\overline{O}_{\overline{m}}(S)\rangle_{S^4}=G_{n\overline{m}}^{S^4}=\frac{1}{\mathcal{Z}(\tau_n,\overline{\tau}_{\overline{n}})} \partial_{\tau_n}\partial_{\overline{\tau}_{\overline{m}}}\mathcal{Z}(\tau_n,\overline{\tau}_{\overline{n}})\,.
\end{equation}

Naively one can suspect that the sphere correlator $\langle O_n(N)\overline{O}_{\overline{m}}(S)\rangle_{S^4}$ can be directly related to the $\mathbb{R}^4$ correlator. However, due to the conformal anomaly, operators in the $S^4$ mix with operators of dimensions lowered in steps of 2 through the curvature schematically as\footnote{We will keep  $S^4,\ \mathbb{R}^4$ superscripts in operators to remind whether we refer to operators on the $S^4$ --which include mixings among different dimensions-- or to operators on $\mathbb{R}^4$ where such mixings are absent. Up to coefficients, the $\mathbb{R}^4$ operators are obtained by orthogonalization of the $S^4$ operators with respect to the inner product defined by the matrix of second derivatives of $\mathcal{Z}$.}

\begin{equation}
\label{mixings}
O^{\mathbb{R}^4}\rightarrow O^{S^4}_{\Delta}+\alpha_1 R\,  O^{S^4}_{\Delta-2}+\alpha_2 R^2\,  O^{S^4}_{\Delta-4}+\cdots
\end{equation}
where $R$ is the Ricci scalar. As a result, the sphere correlation functions lead to mixings which are not expected on $\mathbb{R}^4$ --where the correlators should be of the form \eqref{R4}. The prescription in \cite{Gerchkovitz:2016gxx}  amounts to perform a Gram-Schmidt orthogonalization of the $S^4$ correlators, which then are identified with \eqref{R4}. 

In general, the set of CPO's contains both single-trace and multi-trace operators. For instance, consider superconformal SQCD with gauge group $U(N)$. The CPO's correspond to operators of $O_{n_1\cdots n_m}=({\rm Tr}\phi)^{n_1}\cdots ({\rm Tr}\phi^N)^{n_N}$, where $\phi$ is the scalar in the vector multiplet. Clearly, at a given dimension $\Delta$, the set of CPO's are those satisfying $\sum_m m \,n_m=\Delta$. Then, correlators on the $S^4$ of these operators are given by

\begin{equation}
\langle O^{S^4}_{n_1\cdots n_N}(N) \overline{O}^{S^4}_{\overline{m_1}\cdots \overline{m_N}}(S)\rangle = \frac{1}{\mathcal{Z}} \partial^{n_1}_{\tau_1}\cdots \partial^{n_N}_{\tau_N}\partial^{n_1}_{\overline{\tau}_{\overline{1}}}\cdots \partial^{n_N}_{\overline{\tau}_{\overline{N}}}\mathcal{Z}\Bigg|_{\{\tau_{k\ne 2}\}=\{\overline{\tau}_{\overline{k}\ne 2}\}=0}\, .
\end{equation}
In general there will be mixings between different operators. The Gram-Schmidt procedure amounts to disentangle these mixings by finding an orthogonal basis. It is clear that when constructing such basis, there will be mixings between multi-trace and single-trace operators.

In the following we will be interested on computing correlation functions of operators in gauge theories in the large $N$ limit. In this limit, the mixing between multi- and single-trace operators is suppressed in $N$, and thus we can simply consider single-trace operators (see appendix \ref{DecouplingOfMultitrace} for a review of this feature  in this context). Hence, the relevant set of operators in the large $N$ limit are operators of the form ${\rm Tr}\phi^n$. Note in particular that at each dimension there is one single operator, as opposed to the finite $N$ case, where at each dimension there will be generically a plethora of operators as described above. However, due to the conformal anomaly, there will still be mixings between, say, ${\rm Tr}\phi^n$ and ${\rm Tr}\phi^{n-2}$ which  need to be disentangled through the Gram-Schmidt procedure.

Since in the large $N$ limit we can consider only single-trace operators, the relevant derivatives are

\begin{equation}
\langle O^{S^4}_{n}(N) \overline{O}^{S^4}_{\overline{m}}(S)\rangle = \frac{1}{\mathcal{Z}} \partial_{\tau_n}\partial_{\overline{\tau}_{\overline{m}}}\mathcal{Z}\Bigg|_{\{\tau_{k\ne 2}\}=\{\overline{\tau}_{\overline{k}\ne 2}\}=0}\, .
\end{equation}
Defining as usual $\mathcal{Z}(\tau_k,\overline{\tau}_{\overline{k}})=e^{-\mathcal{F}(\tau_k,\overline{\tau}_{\overline{k}})}$, we then have

\begin{equation}
\langle O^{S^4}_{n}(N) \overline{O}^{S^4}_{\overline{m}}(S)\rangle =  \partial_{\tau_n}\mathcal{F}\partial_{\bar{\tau}_m}\mathcal{F}-\partial_{\tau_n}\partial_{\overline{\tau}_{\overline{m}}}\mathcal{F}\Bigg|_{\{\tau_{k\ne2}\}=\{\overline{\tau}_{\overline{k}\ne2}\}=0}\, .
\end{equation}
Note that $\partial_{\tau_n}\mathcal{F}$ is nothing but the VEV of the operator $O^{S^4}_n$ --which is generically non-zero when the theory is on  $S^4$. Thus, we can redefine a set of zero-VEV  operators $O^{S^4}_n$ by substracting the VEV. Note that we are interested in performing the Gram-Schmidt process at the end of the day, and so the mixing with the identity --VEV-- is one of the particular components that  need to be disentangled. Hence going to these VEV-less operators amounts to a first step in that direction. Moreover, we have that the correlator of these VEV-less operators is

\begin{equation}
\label{VEVlesscorrelatorS4}
\langle O^{S^4}_{n}(N) \overline{O}^{S^4}_{\overline{m}}(S)\rangle = -\partial_{\tau_n}\partial_{\overline{\tau}_{\overline{m}}}\mathcal{F}\Bigg|_{\{\tau_{k\ne 2}\}=\{\overline{\tau}_{\overline{k}\ne 2}\}=0}\, .
\end{equation}
Thus, in the following we will be interested in computing the matrix of second derivatives of $\mathcal{F}$ in different theories, as it is this matrix for which a Gram-Schmidt orthogonalization will lead to the desired correlators in $\mathbb{R}^4$. A more precise description of the orthogonalization procedure is given in appendix \ref{DecouplingOfMultitrace}.

%%%%%%%%%%%%%%%%%
\section{Correlation functions in $\mathcal{N}=4$ at large $N$}\label{N=4}

The matrix model for $\mathcal{N}=4$ SYM on $S^4$ was computed, using supersymmetric localization, in \cite{Pestun:2007rz}, confirming the conjectured matrix model of \cite{Drukker:2000rr}. Introducing the deformations, we can succinctly write it as \cite{Gerchkovitz:2016gxx}

\be
\label{matrixmodelN=4}
\mathcal{Z}=\int d^{N}a\,\Delta(a)\, \Big| e^{i\sum_{n=1}^N\pi^{n/2}\,\tau_n\, \sum_i(a_i)^n }\Big|^2\, ,
\ee
where $\tau_2=\tau_{\rm YM}$. Moreover

\be
\label{vandermonde}
 \Delta(a)=\prod_{i<j}(a_i-a_j)^2\, .
\end{equation}
It is important to stress that the integral over the Cartan of the gauge group in \eqref{matrixmodelN=4} and the measure \eqref{vandermonde} assume that the gauge group is $U(N)$. As such, in addition to the deformations in \cite{Gerchkovitz:2016gxx} corresponding to ${\rm Tr}\phi^n$ for $n\geq 2$, we can add yet one more corresponding to the operator ${\rm Tr}\phi$. 

In the large $N$ limit, the integral \eqref{matrixmodelN=4} can be computed by the saddle-point method (see \textit{e.g.} \cite{Marino:2011nm} for a review). We write

\be
\label{S}
\mathcal{Z}=\int d^{N}a\,e^{-S}\ ,\qquad S=-i\sum_{n=1}^N\pi^{n/2}\,(\tau_n-\bar{\tau}_n)\,\sum_ia_i^n-\sum_{i<j}\log(a_i-a_j)^2\ .
\ee
We also define 't Hooft coupling parameters $\mathfrak{g}_n$

\begin{equation}
\label{calg}
\mathfrak{g}_n=N^{-1}\pi^{\frac{n}{2}}\tau_n\, .
\end{equation}
Note that the action depends on these through

\begin{equation}
\label{hhgg}
g_n=2\,{\rm Im}\mathfrak{g}_n,
\end{equation}
The large $N$ limit is defined by taking the limit $N\to\infty $ with fixed $\mathfrak{g}_n$. Since these are the natural variables to use in the large $N$ limit, we will adapt the prescription in \cite{Gerchkovitz:2016gxx}  whereby we consider derivatives of the deformed partition function with respect to $\mathfrak{g}_n,\,\bar{\mathfrak{g}}_n$ instead of with respect to $\tau_n,\,\bar{\tau}_n$. Of course, at the end we need to set to zero all $\mathfrak{g}_n$ but $\mathfrak{g}_2\sim \tau_{YM}$. Note that we reabsorb in the $\mathfrak{g}_n$'s a factor of $\pi^{n/2}$ present in the deformed matrix model of eq. \eqref{matrixmodelN=4}. Thus, derivatives with respect to $\mathfrak{g}_n$ insert the operator ${\rm Tr}\phi^n$, whose correlators in $\mathbb{R}^4$ we will be computing.

Introducing as usual the eigenvalue density normalized to one

\begin{equation}
\rho(x)=\frac{1}{N}\sum_i\delta(x-a_i)\ ,\qquad \int dx\ \rho(x)=1\ ,
\end{equation}
one can see that $\rho$ is determined from the singular integral equation

\begin{equation}
\label{eom2}
\dashint dy\frac{\rho(y)}{x-y}=\frac{1}{2}V'(x)\, ,\qquad V(x)=\sum_{n=1}^N g_n\,x^n\, .
\end{equation}
\begin{equation}
\label{F}
 N^{-2} S=F=\int dx \, \rho(x)V(x)- \frac{1}{2} \int dx\int dy \rho(x)\rho(y)\ln (x-y)^2 \, .
\end{equation}
In the large $N$ limit the value of $\mathcal{Z}$ is given by the saddle point approximation

\begin{equation}
\mathcal{Z}=e^{-\mathcal{F}(\tau_k,\overline{\tau}_k)}\rightarrow e^{-N^2 F|_{\rm saddle}}\, ,
\end{equation}
being $F|_{\rm saddle}$ given by the evaluation of \eqref{F} on the $\rho$ arising from solving \eqref{eom2}. 

The solution of the matrix model  is known for a general potential. 
For a polynomial potential, assuming that eigenvalues  condense in a cut $(-\nu,\mu )$, it is of the form
\be
\rho(x)= \left( \sum_{k=0}^{n_1} c_k x^k \right) \sqrt{(\mu-x)(x +\nu)}\, ,
\ee 
where $n_1$ depends on the degree of the polynomial.
The matrix model can have multicut solutions appearing for critical values of the couplings.
However,  multicut solutions are not relevant for the present discussion, because here we assume that all
deformations, $\mathfrak{g}_n$ with $n\neq 2$, are small. In this case, the eigenvalue distribution is a mild deformation of the Wigner semi-circle distribution that describes the large $N$ limit of ${\cal N}=4$ super Yang-Mills theory. 

Before setting the $\{\mathfrak{g}_{i\ne2}\}$ to zero the cut will generically be asymmetric in the real line due to the presence of odd terms in $x$ in the potential $V$. In particular, the condition $\sum_{i=1}^Na_i=0$, which in the continuum becomes $\int_{-\nu}^{\mu}dx\,\rho(x)\,x=0$, will not be satisfied, explicitly manifesting that the computation is for a $U(N)$ theory (the $SU(N)$ theory is discussed in section 3.2).

\subsection{Correlation functions in the matrix model}

\subsubsection*{Deformation by even operators}

Even and odd operators do not mix in correlation functions in ${\cal N}=4$ theory. A direct way to see this is to note that
the integral
\be
\int d^{N}a\,\Delta(a)\, \ |e^{i \pi\,\tau\,\sum_i a_i^2}|^2\ (\sum_k a_k^n) (\sum_j a_j^m)  
\ee
vanishes if $n$ and $m$  have distinct parity, since all $a_i$ are integrated from $(-\infty,\infty)$.
Thus, for clarity in the presentation, we may begin our discussion by restricting the consideration to the ${\cal N}=4$ theory deformed by even operators, in which case the matrix model has reflection symmetry leading to $\nu=\mu$ and the solution is much simpler. Note that as a consequence of the reflection symmetry the condition $\int dx\,\rho(x) \,x=0$ will be met, and thus the calculations in this sector also
apply to the $SU(N)$ theory.

Since for now we are including only even deformations, it is useful to redefine the potential as

\be
V= \sum_{n=1}^{n_0} g_{2n} x^{2n}\ ,\ \  \ \ \ n_0\equiv [N/2]\ .
\ee
Since the potential is invariant under reflection symmetry, we can assume that eigenvalues
will condense in a  cut $(-\mu,\mu)$, i.e. in this case $\nu =\mu$.

By acting with 
\be
\dashint_{-\mu}^\mu dx \frac{1}{\sqrt{\mu^2-x^2}}\ \frac{1}{z-x}\ 
\ee
on (\ref{eom2}), we get
\be
\rho(z) =\frac{1}{4\pi^2} \sqrt{\mu^2-z^2} \dashint_{-\mu}^\mu dx \frac{V'(x)}{\sqrt{\mu^2-x^2}(z-x)}\, ,
\ee
i.e.
\be
\rho(z) =\frac{1}{2\pi^2} \sqrt{\mu^2-z^2} \sum_{n=1}^{n_0}n g_{2n}  \dashint_{-\mu}^\mu dx \frac{x^{2n-1}}{\sqrt{\mu^2-x^2}(z-x)}\, .
\ee
The integral can be computed by choosing a contour that surrounds the cut $(-\mu,\mu)$ and computing the residue at infinity as

\be
\label{i1}
 \dashint_{-\mu}^\mu dx \frac{x^{2n-1}}{\sqrt{\mu^2-x^2}(z-x)}=2\pi \sum_{k=0}^{n-1} b_k z^{2n-2k-2} \mu^{2k}\, .
\ee
Then, we obtain
\be
\rho(x) =\left(\sum_{k=0}^{n_0-1} q_k z^{2k}\right) \sqrt{\mu^2-z^2}\,,
\ee
with
\be
\label{qs}
q_k=\frac{1}{\pi} \sum_{n=k+1}^{n_0} n  b_{n-k-1} g_{2n} \mu^{2n-2k-2}\ ,\ \ \
\ee
\be
 b_k\equiv \frac{1}{\sqrt{\pi}} \frac{ \Gamma(k+1/2)}{k!}\, .
\ee
Let us now consider the normalization condition.
Again, by residues
\be
\int_{-\mu}^\mu dz\ z^{2k} \sqrt{\mu^2-z^2} = \pi \aaa_{k+1} \mu^{2k+2}\ ,
\ee
where
\be
\aaa_k \equiv \frac{1}{2\sqrt{\pi}} \frac{\Gamma(k-1/2)}{k!}\, .
\ee
Hence, normalization implies
\be
\pi \sum_{k=0}^{n_0-1} q_k \aaa_{k+1} \mu^{2k+2} =1\, .
\ee
Using the expression for $c_k$, this becomes
\be
\label{normapar}
\sum_{n=1}^{n_0}   n b_n g_{2n} \mu^{2n} =1\, .
\ee
where we used the identity
\be
 \sum_{k=0}^{n-1}  \aaa_{k+1} b_{n-k-1} = b_n\, .
\ee

As described above, in order to compute the $\mathbb{R}^4$ correlators we are interested on, we need to compute the $\mathfrak{g}_n$ derivatives of $\mathcal{F}$. Since $g_n=2{\rm Im}\mathfrak{g}_n=i(\bar{\mathfrak{g}}_n-\mathfrak{g}_n)$, we have that $\partial_{\mathfrak{g}_n}\partial_{\bar{\mathfrak{g}}_m}=\partial_{g_n}\partial_{g_m}$, and thus the derivatives of interest  coincide with those with respect to $g_n$. Thus, in order to compute two point functions in large $N$ we need to compute the matrix of second derivatives of $F$ as in eq. \eqref{VEVlesscorrelatorS4}. To that matter, we begin with the formula

\be
\partial_{g_{2n}} F= \int_{-\mu}^\mu dz \ z^{2n} \rho(z) \, .
\ee
This gives
\be
\partial_{g_{2n}} F = \pi \sum_{k=0}^{n_0-1} q_k \aaa_{k+n+1}  \mu^{2k+2n+2} = \sum_{m=1}^{n_0} d_{m,n} g_{2m} \mu^{2m+2n} \, ,
\ee
with
\be
\label{ddmn}
d_{m,n} \equiv m\sum_{k=0}^{m-1}  b_{m-k-1} \aaa_{k+n+1} =\frac{(2m)! \Gamma(n+\frac{1}{2})}{4^m\sqrt{\pi}(m+n) n! (m-1)!^2}\, .
\ee

Next, we compute the second derivative of the free energy, 
\be
\partial_{g_{2m}}\partial_{g_{2n}} F = \sum_{k=1}^{n_0}  (2k+2n) d_{k,n} g_{2k} \mu^{2k+2n-1} \frac{d\mu}{dg_{2m}}+ d_{m,n} \mu^{2m+2n} \, .
\ee
To compute
 $d\mu/dg_{2m}$, we use  the normalization condition.  This is done in appendix \ref{details}.
After differentiation, we
must set all $g_{2n}=0$, with $n>1$. We get
\be
\frac{d\mu}{dg_{2k}}= -\frac{k b_{k}\mu^{2k-1}}{  g_{2}} =-\frac{1}{2}kb_k \mu^{2k+1}\ ,\
\ee
where we used
\be
 \mu ^2= \frac{2}{g_2} =\frac{\lambda}{(2\pi)^2} \, .
\ee
Therefore
\be\label{desp}
\partial_{g_{2m}}\partial_{g_{2n}}  F\bigg|_{g_{2k>2}=0} = \big(d_{m,n} - (2+2n) d_{1,n}m b_m\big) \mu^{2m+2n}\, .
\ee
After some simple algebra, we obtain
\be\label{eveneven}
\partial_{g_{2m}}\partial_{g_{2n}} F = - \left(\frac{\lambda}{4\pi^2}  \right)^ {m+n} \frac{\Gamma(m+\frac{1}{2})\Gamma(n+\frac{1}{2}) }{\pi (m+n) \Gamma(m) \Gamma(n) }\, .
\ee

In appendix \ref{details} we also give the expressions of $\partial_{g_{2\ell } }\partial_{g_{2j}}\partial_{g_{2n}} F $ and
$\partial_{g_{2s}}\partial_{g_{2k}} \partial_{g_{2j}}\partial_{g_{2n}} F $.

\subsubsection*{General Deformation including even and odd operators}

Let us now apply the methods developed for the even deformations to the more involved case where we allow both even and odd powers in the potential. In that case the eigenvalues will not be distributed symmetrically around the origin. Nevertheless, we can follow the same logic by acting on the integral equation (\ref{eom2}) with 
\be
\dashint_{-\nu}^\mu dx \frac{1}{\sqrt{(\mu-x)(x +\nu)}}\ \frac{1}{z-x}\ .
\ee
We get
\be
\rho(z) =\frac{1}{4\pi^2} \sqrt{(\mu-z)(z +\nu)} \dashint_{-\nu}^\mu dx \frac{V'(x)}{\sqrt{(\mu-x)(x +\nu)}(z-x)}\ ,
\ee
i.e.
\be
\rho(z) =\frac{1}{4\pi^2} \sqrt{(\mu-z)(z +\nu)}\  \sum_{n=1}^{N}n  g_n  \dashint_{-\nu}^\mu dx \frac{x^{n-1}}{\sqrt{(\mu-x)(x +\nu)}(z-x)}\ .
\ee
Computing this integral by residues (using a contour surrounding the cut $(-\nu,\mu)$),
we get   
\be
c_k = 
\frac{1}{2\pi} \sum_{n=k+2}^N n g_n \sum_{r =0}^{n-k-2}    b_{r} b_{n-k -r-2}   \mu^{r}(-\nu )^{n-k -r-2}\ ,
\ee
and
\be
\rho(x)= \left( \sum_{k=0}^{N-2} c_k x^k \right) \sqrt{(\mu-x)(x +\nu)}\, .
\ee

In order to compute the two-point functions of interest we need to compute the matrix of second derivatives of $F$ with respect to the couplings $g_n$. As before we start with 
\be
\partial_{g_\ell } F= \int_{-\nu}^\mu dz \ z^{\ell } \rho(z) \equiv m_{\ell } \ .
\ee
Using the explicit form of the density, we obtain for the moments $m_{\ell}$, 
\be
m_{\ell } 
= -\pi \sum_{k=0}^{N-2} c_k\sum_{r=0}^{k+\ell+2}
\aaa_r \aaa_{k+\ell+2-r} \mu^r (-\nu)^{k+\ell+2-r}\,.
\ee

We now need to differentiate with respect to $g_s$. Recall that eventually we want to evaluate these derivatives upon setting all $g_{n\ne 2}$ to zero. This also implies
setting, after differentiation, all $c_k=0,\ k>0$, $c_0=g_2/\pi$ and $\mu=\nu $.
One contribution comes from
\bea
\partial_{g_s } c_k &=& \frac{s}{2\pi} \sum_{r=0}^{s-k-2} b_r b_{s-k-r-2}\mu^r (-\nu)^{s-k-r-2} \\ 
&=& \frac{s}{2\pi} (-1)^{s-k} \mu^{s-k-2} \sum_{r=0}^{s-k-2} (-1)^r b_r b_{s-k-r-2}
\nonumber\\
 &=& \frac{s}{2\pi}\mu^{s-k-2}  (-1)^{s-k}\gamma_{s-k-2}\  \theta(s-k-2)\, ;
\label{cuarenta}
\eea
with
\be\label{gamita}
\gamma_{m}=
(1+(-1)^{m}) \frac{\Gamma(\frac{m}{2} +\frac{1}{2})}
{2\sqrt{\pi} \Gamma(1+\frac{m}{2} )}\,,
\ee
and $\theta(x) $ is the step function (with the convention $\theta(0)=1$). Note that $\gamma_{2n}=b_n$.

Other contributions come from $\partial_{g_s }\mu,\ \partial_{g_s }\nu$. 
These are computed in appendix \ref{details}. We obtain
\bea
&& g_2 \partial_{g_s} \mu = g_2 \partial_{g_s} \nu = -\frac{1}{2}\mu^{s-1} s b_{\frac{s}{2}}\, ,\qquad \ \quad {\rm for\ even} \ s\, ;
\nonumber\\
&&
g_2\partial_{g_s} \mu =- g_2\partial_{g_s} \nu =-\frac{1}{2}\mu^{s-1}s\gamma_{s-1}\, ,
\qquad {\rm for\ odd} \ s\, .
\eea
Using the above formulas, we find
\bea
\partial_{g_s }\partial_{g_\ell } F\bigg|_{g_{n\neq  2}=0} &=& -\pi
\sum_{k=0}^{s-2} \partial_{g_s} c_k (-1)^{k+\ell} \mu^{k+\ell+2} \alpha_{k+\ell+2}
- g_2 (\ell+2) (-1)^{\ell } \mu^{\ell+1}  \alpha_{\ell+2} \partial_{g_s} \nu
\nonumber\\
&-& g_2 (-1)^{\ell } \mu^{\ell+1} \beta_{\ell+2} ( \partial_{g_s} \mu- \partial_{g_s} \nu)\, ;
\eea
where

\bea
\alpha_m&\equiv &\sum_{r=0}^{m} (-1)^r \aaa_r \aaa_{m-r} =- (1+(-1)^m) \frac{\Gamma(\frac{m}{2} -\frac{1}{2})}
{4\sqrt{\pi} \Gamma(\frac{m}{2}+1 )}\, ;
\nonumber\\
\beta_m &\equiv&  \sum_{r=0}^{m} (-1)^r r \aaa_r \aaa_{m-r}
=\begin{cases}
 -  \frac{\Gamma(\frac{m}{2} -\frac{1}{2})}
{2\sqrt{\pi} \Gamma(\frac{m}{2} )}\ ,\qquad m\ {\rm even}\, ; \\
\\
  \frac{\Gamma(\frac{m+1}{2} -\frac{1}{2})}
{2\sqrt{\pi} \Gamma(\frac{m+1}{2} )}\ ,\qquad m\ {\rm odd}\, .
\end{cases}
\eea
Note that $\alpha_{2k}=-\aaa_k$, $\beta_{2k}=- k \aaa_k =- \frac{1}{2} b_{k-1},\ 
\beta_{2k+1}= \frac{1}{2}b_k $.

Thus
\bea\label{ccro}
\partial_{g_s }\partial_{g_\ell } F\bigg|_{g_{n\neq 2}=0} &=& -\frac{1}{2} (-1)^{s+\ell} \mu^{\ell+s}
\ s \sum_{k=0}^{s-2}   \alpha_{k+\ell+2}  \gamma_{s-k-2}
- g_2 (\ell+2) (-1)^{\ell } \mu^{\ell+1}  \alpha_{\ell+2} \partial_{g_s} \nu
\nonumber\\
&-& g_2 (-1)^{\ell } \mu^{\ell+1} \beta_{\ell+2} ( \partial_{g_s} \mu- \partial_{g_s} \nu)\, .
\eea

It is now easy to demonstrate that non-vanishing two-point functions $\partial_{g_s }\partial_{g_\ell } F$ have both $s$ and $\ell $ of the same parity.
Indeed, if $\ell $ is odd and $s $ is even (or viceversa) the first term in (\ref{ccro}) vanishes,
because  the product $\alpha_{k+\ell+2}  \gamma_{s-k-2}$ only contributes where the arguments of both $\alpha_m$ and $\gamma_m$ are even.
The second term and third term also vanish when $\ell $ is odd and $s $ is even.
In the other case, $\ell $ even and $s$ odd, the second and third term become
$$
 g_2 \mu^{\ell+1} \partial_{g_s} \mu \big((\ell+2) \alpha_{\ell+2} -2 \beta_{\ell+2} \big) =0\, ,
$$
where we used that $m\alpha_m=2\beta_m$ for even $m$.

Thus we need to consider two cases, 1) $s,\ \ell$ even, and 2) $s,\ \ell $ odd.

\begin{itemize}

\item Even-Even two-point function: Let $s=2m$, $\ell =2n$.
Then
\bea
2m \sum_{k=0}^{2m-2}   \alpha_{k+2n+2}  \gamma_{2m-k-2} &=& 2m
\sum_{r=0}^{m-1}   \alpha_{2r+2n+2}  \gamma_{2m-2r-2}
\nonumber\\
&=& -2m \sum_{r=0}^{m-1}   \aaa_{r+n+1}  b_{m-r-1}= - 2 d_{m,n}\, ,
\eea
and
\be\label{ccop}
\partial_{g_{2m}}\partial_{g_{2n} } F =  \mu^{2m+2n}\Big(
d_{m,n}
- m b_n b_m\Big)\ .
\ee
This  simplifies to\footnote{The equivalence between (\ref{ccop}) and (\ref{desp})
follows from the simple identities $d_{1,n}=\aaa_{n+1}$ and $b_n=2(1+n) \aaa_{n+1}$.}

\be\label{eveneven2}
\partial_{2m}\partial_{2n} F = -  \left( \frac{\lambda}{4\pi^2}  \right)^ {m+n} \frac{\Gamma(m+\frac{1}{2})\Gamma(n+\frac{1}{2}) }{\pi (m+n) \Gamma(m) \Gamma(n) }\, ,
\ee
which reproduces (\ref{eveneven}).

\item Odd-Odd two-point function: Let $s=2m+1$, $\ell =2n+1$.
Now
\bea
(2m+1) \sum_{k=0}^{2m-1}   \alpha_{k+2n+3}  \gamma_{2m-k-1} &=& (2m+1)
\sum_{r=0}^{m-1}   \alpha_{2r+2n+4}  \gamma_{2m-2r-2}
\nonumber\\
&=& -(2m+1) \sum_{r=0}^{m-1}   \aaa_{r+n+2}  b_{m-r-1}
\nonumber\\
&=& - \frac{2m+1}{m} d_{m,n+1}\, .
\eea
Therefore
\be\label{ccor}
\partial_{g_{2m+1}}\partial_{g_{2n+1} } F =  \mu^{2m+2n+2} \frac{2m+1}{2m} \Big(  d_{m,n+1}
-   m b_{m}  b_{n+1}  \Big)\, .
\ee
Substituting the expressions for the coefficients $d_{m,n+1}, b_m$,
we finally find
\be\label{finalodd}
\partial_{g_{2m+1}}\partial_{g_{2n+1} } F = - \Big(\frac{\lambda}{4\pi^2} \Big)^{m+n+1} \frac{\Gamma\big(m+\frac{3}{2} \big)\Gamma\big(n+\frac{3}{2} \big)}{
\pi (m+n+1) \Gamma(m+1) \Gamma(n+1)}\ .
\ee

\end{itemize}

\vspace{1cm}

\subsubsection*{Gram-Schmidt orthogonalization}

We now need to run the Gram-Schmidt orthogonalization procedure (further details can be seen in appendix \ref{DecouplingOfMultitrace}). As discussed above, in the large $N$ limit it is more natural to consider the $\mathfrak{g}_n$ variables. Then, orthogonalization of the matrix of $\mathfrak{g}$-derivatives, and upon taking into account a factor of $4^{n}$ as in \cite{Gerchkovitz:2016gxx}\footnote{The reason for this can be traced to the conformal mapping from $\mathbb{R}^4$ to $S^4$. Consider the $\mathbb{R}^4$ operator $\lim_{x\rightarrow \infty}x^{2\Delta}O_{\Delta}(x)$ and re-write it as $4^{\Delta}\lim_{x\rightarrow \infty}\Big(\frac{x^2}{4}\Big)^{\Delta}O_{\Delta}(x)\sim 4^{\Delta}\lim_{x\rightarrow \infty}\Big(1+\frac{x^2}{4}\Big)^{\Delta}O_{\Delta}(x)$. Since the conformal mapping $\mathbb{R}^4$ into $S^4$ is $ds^2_{\mathbb{R}^4}=(1+\frac{\vec{x}^2}{4})^2ds^2_{S^4}$, this is simply  $4^{\Delta}O(N)$. Thus $\langle O^{\mathbb{R}^4}(0)\overline{O}^{\mathbb{R}^4}(0)\rangle_{\mathbb{R}^4} = 4^{\Delta}\,\langle O^{S^4}(S)\overline{O}^{S^4}(N)\rangle_{S^4}$.} , computes the correlators of ${\rm Tr}\phi^n$ in $\mathbb{R}^4$. 

Since the mixed odd-even derivatives of $F$ vanish, we can run this procedure in the even and odd sectors separately. It is straightforward to check that the first few orthogonalized operators  are (recall $O^{S^4}_n$ refers to VEV-less operators on the $S^4$)

\begin{eqnarray}
\label{OperatorsN=4}
&&O^{S^4}_1\,; \nonumber \\
&&O^{S^4}_2\,;\nonumber \\
&& O^{S^4}_3-\frac{3\lambda}{(4\pi)^2}O^{S^4}_1\,; \nonumber\\
&&O^{S^4}_4-\frac{4\lambda}{(4\pi)^2} O^{S^4}_2\,;\nonumber \\
&&O^{S^4}_5-\frac{5\lambda}{(4\pi)^2}O^{S^4}_3+\frac{5\lambda^2}{(4\pi)^4}O^{S^4}_1\,;\nonumber \\
&&O^{S^4}_6-\frac{6\lambda}{(4\pi)^2} O^{S^4}_4+\frac{9\lambda^2}{(4\pi)^4}O^{S^4}_2\,;\nonumber \\
&&\cdots
\end{eqnarray}
Taking into account the numerical coefficient explained above, for $O^{\mathbb{R}^4}_n={\rm Tr}\phi^n$ on $\mathbb{R}^4$, we finally obtain

\begin{equation}
\label{N=4CPO}
\langle O^{\mathbb{R}^4}_n(0)\overline{O}^{\mathbb{R}^4}_{\overline{m}}(x)\rangle_{\mathbb{R}^4}=\frac{\delta_{n\overline{m}}}{|x|^{2\Delta_n}}\frac{\Delta_n\,\lambda^{\Delta_n}}{(2\pi)^{2\Delta_n}}\, .
\end{equation}
This exactly coincides with the result in \cite{Lee:1998bxa} computed in the free theory, which, as a consequence of a non-renormalization theorem \cite{Lee:1998bxa,Freedman:1998tz,D'Hoker:1998tz,Penati:1999ba,Baggio:2012rr}, holds to all loop orders (up to ambiguous contact terms).

\subsection{The $SU(N)$ theory}

So far we have concentrated on the $U(N)$ theory. Let us now consider the $SU(N)$ theory, which amounts to demand that $\sum_{i=1}^N a_i =0 $ in eq. \eqref{matrixmodelN=4}. This can be implemented by inserting in the integral $\delta(\sum_{i=1}^N a_i)$. Writing the $\delta$ in Fourier space, it becomes evident that the momentum variable of the integration appears just like $\tau_1$ in eq.\eqref{S}. Thus, we can recover the $SU(N)$ case by simply integrating over $\tau_1$, since $\tau_1$ is playing the role of Lagrange  multiplier enforcing the tracelessness condition of $SU(N)$ (for simplicity of the presentation, we reabsorb all factors of $\pi$ in the couplings). Since odd-even mixed derivatives do not couple, we can consider for this matter just the odd deformations. Consistently, as described above, the even deformations involve solutions where the cut where eigenvalues live is symmetric in the real axis around the origin, and thus the even correlation functions directly coincide in $SU(N)$ and $U(N)$. Moreover, note that as a consequence of the fact that even-odd derivatives vanish the VEV's of odd operators vanish (as it should be expected on general grounds from eq. \eqref{mixings}). Thus for odd operators $O_n^{S^4}={\rm Tr}\phi^n$. Then we can write

\begin{equation}
\mathcal{F}=\langle {\rm Tr}\phi^{2n+1}{\rm Tr}\overline{\phi}^{2m+1}\rangle \tau_{2n+1}\bar{\tau}_{2m+1}+\mathcal{F}_{\rm even}\,, \qquad n,\,m\geq 0\,;
\end{equation}
where $\langle {\rm Tr}\phi^{2n+1}{\rm Tr}\overline{\phi}^{2m+1}\rangle$ stands for the $S^4$ correlators arising from the matrix of second derivatives in eq.\eqref{finalodd}. For simplicity of the presentation, let us consider operators up to dimension $\Delta=3$. Then, the relevant part of the partition function is

\begin{equation}
\mathcal{Z}=e^{\langle {\rm Tr}\phi{\rm Tr}\overline{\phi}\rangle \tau_{1}\bar{\tau}_{1}+\langle {\rm Tr}\phi{\rm Tr}\overline{\phi}^{3}\rangle \tau_{1}\bar{\tau}_{3}+\langle {\rm Tr}\phi^{3}{\rm Tr}\overline{\phi}\rangle\tau_{3}\bar{\tau}_{1}+\langle {\rm Tr}\phi^{3}{\rm Tr}\overline{\phi}^{3}\rangle \tau_{3}\bar{\tau}_{3}+\cdots}\, .
\end{equation}
This can be re-written as

\begin{equation}
\mathcal{Z}=e^{ \langle {\rm Tr}\phi{\rm Tr}\overline{\phi}\rangle \hat{\tau}_1\overline{\hat{\tau}}_1 + \big(\langle {\rm Tr}\phi^3{\rm Tr}\overline{\phi}^3\rangle -\frac{\langle {\rm Tr}\phi {\rm Tr}\overline{\phi}^3\rangle \langle {\rm Tr}\phi^3{\rm Tr}\overline{\phi}\rangle}{\langle{\rm Tr}\phi{\rm Tr}\overline{\phi}\rangle}\big)\tau_3\overline{\tau}_3+\cdots}\, ,
\end{equation}
where

\begin{equation}
\hat{\tau}_1=\tau_1-\frac{\langle {\rm Tr}\phi{\rm Tr}\overline{\phi}^3\rangle}{\langle{\rm Tr}\phi{\rm Tr}\overline{\phi}\rangle}\tau_3\,.
\end{equation}
Thus, the $SU(N)$ theory is obtained by integrating over $\tau_1$. Performing the shift of the integration into $\hat{\tau}_1$, the result is simply

\begin{equation}
\label{N=4SU(N)}
\mathcal{Z}_{SU(N)}=e^{\big(\langle {\rm Tr}\phi^3{\rm Tr}\overline{\phi}^3\rangle -\frac{\langle {\rm Tr}\phi {\rm Tr}\overline{\phi}^3\rangle \langle {\rm Tr}\phi^3{\rm Tr}\overline{\phi}\rangle}{\langle{\rm Tr}\phi{\rm Tr}\overline{\phi}\rangle}\big)\tau_3\overline{\tau}_3+\cdots}\, .
\end{equation}
Inspection of eq.\eqref{N=4SU(N)} shows that the mixing with ${\rm Tr}\phi$ is gone --as this operator is absent in the $SU(N)$ theory--, and thus ${\rm Tr}\phi^3$ does not mix with any operator. Thus, we can easily read the $\mathbb{R}^4$ correlator for ${\rm Tr}\phi^3$ from $\mathcal{Z}_{SU(N)}^{-1}\partial_{\tau_3}\partial_{\overline{\tau}_3}\mathcal{Z}_{SU(N)}$

\begin{equation}
\langle {\rm Tr}\phi^3{\rm Tr}\overline{\phi}^3\rangle_{\mathbb{R}^4}=\frac{1}{|x|^6}\Big(\langle {\rm Tr}\phi^3{\rm Tr}\overline{\phi}^3\rangle_{S^4} -\frac{\langle {\rm Tr}\phi {\rm Tr}\overline{\phi}^3\rangle_{S^4} \langle {\rm Tr}\phi^3{\rm Tr}\overline{\phi}\rangle_{S^4}}{\langle{\rm Tr}\phi{\rm Tr}\overline{\phi}\rangle_{S^4}}\Big)=\frac{1}{|x|^6}\frac{3\lambda^3}{(2\pi)^6}\,;
\end{equation}
which is exactly the same result as for the $U(N)$ case. Note that this holds exactly, since

\begin{equation}
\frac{1}{\mathcal{Z}_{SU(N)}}\partial_{\tau_3}\partial_{\overline{\tau}_{\overline{3}}}\mathcal{Z}_{SU(N)}=\langle {\rm Tr}\phi^3{\rm Tr}\overline{\phi}^3\rangle_{S^4} -\frac{\langle {\rm Tr}\phi {\rm Tr}\overline{\phi}^3\rangle_{S^4} \langle {\rm Tr}\phi^3{\rm Tr}\overline{\phi}\rangle_{S^4}}{\langle{\rm Tr}\phi{\rm Tr}\overline{\phi}\rangle_{S^4}}
\end{equation}
is exactly the correlator of the $O^{\mathbb{R}^4}_3$ operator (\textit{cf.} eq.\eqref{correlatorSU(N)}). Indeed, the argument above extends straightforwardly to all orders, thus showing the identity of the $SU(N)$ and $U(N)$ results for all operators (obviously aside of ${\rm Tr}\phi$). This is indeed what we should have expected, as the non-renormalization theorem continues to hold, and the difference $SU(N)$ and $U(N)$ for all operators other than ${\rm Tr}\phi$ is subleading in $N$.

\section{Correlation functions in $\mathcal{N}=2$ superconformal QCD at large $N$}\label{sect:SQCD}

The matrix model for $\mathcal{N}=2$ superconfomal QCD was constructed in \cite{Pestun:2007rz}. Then, the deformed partition function is defined as follows:
\be
\label{SCQD}
\mathcal{Z}_{{\cal N}=2\ {\rm SCF}}=
\int d^{N}a\,\Delta(a)\,|e^{i\sum_{n=1}^N\pi^{n/2}\,\tau_n\, \sum_i a_i^n}|^2 
\frac{\prod_{i<j} H^2(a_i-a_j)}{\prod_i H(a_i)^{2N}}
|{\cal Z}_{\rm inst}|^2\, ;
\ee
where
\be
H(x) \equiv\prod_{n=1}^\infty \left( 1+\frac{x^2}{n^2}\right)^n\ e^{-\frac{x^2}{n}}\, .
\ee
The instanton factor $|{\cal Z}_{\rm inst}|\to 1$ exponentially at large $N$, so it will not be considered in what follows. Under that assumption, the remainder of the undeformed matrix model depends  on the squared of the eigenvalues $a_i$. Hence, due to the reflection symmetry $a_i\rightarrow -a_i$ of the measure, correlators mixing even-and odd operators will vanish again. Thus, we can separately treat each case. For simplicity, let us concentrate in the following in the case of even operators.

Just as before, we will be interested in the large $N$ limit, where, due to the same argument as in the case of $\mathcal{N}=4$ SYM, we can concentrate on single-trace operators. Note that in eq.\eqref{SCQD} we are considering the $U(N)$ theory. However, just as in the $\mathcal{N}=4$ case, since the eigenvalue density for only-even deformations is symmetric in the real line around the origin, the solution is identical to that of the $SU(N)$ theory. 

Similarly as before, we define
\be
V= \sum_{n=1}^{n_0} g_{2n} x^{2n}\ ;
\ee
where $g_{2n}=2\,{\rm Im}\mathfrak{g}_n$ are the natural variables in the large $N$ limit and the ones with respect to which we will take derivatives when computing correlators. We then have to solve the saddle-point equation
\be
\label{ferp}
\dashint_{-\mu}^\mu dy\ \rho(y) \left( \frac{1}{x-y}-K(x-y)\right) =\frac{1}{2} V'(x) -K(x)\ ,
\ee
where
\be
K(x)\equiv -\frac{H'(x)}{H(x)}\ .
\ee
This equation was investigated in great detail for the undeformed case in \cite{Passerini:2011fe}.
Following \cite{Passerini:2011fe}, we now get
\be\label{trese}
\rho(z) =\frac{1}{2\pi^2} \sqrt{\mu^2-z^2} \dashint_{-\mu}^\mu dx \frac{1}{\sqrt{\mu^2-x^2}(z-x)} \left(\frac{1}{2}V'(x)
-K(x)+\int dy \rho(y) K(x-y)\right)\, .
\ee

\subsection{Weak coupling}

We aim to find the correlators in the weak coupling regime. This is akin to consider the planar expansion of the theory.
Since $g_2=8\pi^2/\lambda$, at weak coupling, $g_2\gg 1$ and the linear force $V'= 2g_2 x+...$
makes eigenvalues condense near the origin. As a result $\mu<<1$ and we can use the Taylor expansion for the function $K(x)$,
\be
K(x)= -2\sum_{n=1}^\infty (-1)^n \zeta(2n+1) x^{2n+1}\ .
\ee
Using this, the last integral in eq.\eqref{trese} becomes 
\bea
\int_{-\mu}^\mu dy \rho(y) K(x-y) &=& -2\sum_{n=1}^\infty (-1)^n \zeta(2n+1)\int_{-\mu}^\mu  dy \rho(y) (x-y)^{2n+1}
\nonumber\\
 &=& -2\sum_{n=1}^\infty (-1)^n \zeta(2n+1) \sum_{k=0}^{n} {2n+1\choose 2k} 
x^{2n+1-2k} m_{2k}\, ;
\nonumber
\eea
where we have used the fact that the eigenvalue density is even (since the potential and
the one-loop determinant are even) and
\be
m_{2k}\equiv  \int_{-\mu}^\mu  dy\  \rho(y) y^{2k}\,.
\ee
Next, we need to consider the integral which we already computed in eq.\eqref{i1}. Thus we obtain
\be
\rho(x) =\left(\sum_{k=0}^{\infty} C_k z^{2k}\right) \sqrt{\mu^2-z^2}\, ,
\ee
where $C_k$ has three contributions,
\be
C_k=q_k+A_k+B_k
\ee
The first one, $q_k$, is the same as in  ${\cal N}=4$ theory in the even-even case in eq.\eqref{qs} ($q_k=0$ for $k>n_0-1$).
The second one $A_k$ comes from $K(x)$ in eq. (\ref{trese}). The factor multiplying $\sqrt{\mu^2-z^2}$
is
\be
\frac{2}{\pi} \sum_{n=1}^\infty (-1)^n \zeta(2n+1)\sum_{k=0}^{n} b_{n-k} z^{2k} \mu^{2n-2k}
=\sum_{k=0}^\infty A_k z^{2k} \, ,
\ee
with
\be
A_k =\frac{2}{\pi} \sum_{n={\rm max}(k,1)}^\infty (-1)^n \zeta(2n+1) b_{n-k} \mu^{2n-2k}\ ,\qquad k\geq 0\, .
%A_0 &=& \frac{2}{\pi} \sum_{n=1}^\infty (-1)^n \zeta(2n+1) b_{n} \mu^{2n}\, .
\ee
Finally, consider the third contribution $B_k$:
\bea
\sum_{\ell=0}^\infty B_{\ell } z^{2\ell } &=&
-\frac{2}{\pi}\sum_{n=1}^\infty (-1)^n \zeta(2n+1) \sum_{k=0}^{n} {2n+1\choose 2k} 
 m_{2k} \sum_{\ell =0}^{n-k} b_{n-k-\ell} \mu ^{2n-2k-2\ell} z^{2\ell}
\nonumber\\
&=&
-\frac{2}{\pi}\sum_{n=1}^\infty (-1)^n \zeta(2n+1) \sum_{\ell =0}^{n} z^{2\ell} 
  \sum_{k =0}^{n-\ell}  {2n+1\choose 2k} m_{2k} b_{n-k-\ell} \mu ^{2n-2k-2\ell}\, .
\nonumber
\eea
Thus
\be
 B_{\ell} =
-\frac{2}{\pi}
\sum_{n={\rm max}(\ell,1) }^\infty (-1)^n \zeta(2n+1)  \sum_{k =0}^{n-\ell} {2n+1\choose 2k} 
 m_{2k} b_{n-k-\ell} \mu ^{2n-2k-2\ell}\ ,\ \ \ell\geq 0\, .
%\nonumber\\
%&& B_{0} =
% -\frac{2}{\pi}
% \sum_{n=1 }^\infty (-1)^n \zeta(2n+1)  \sum_{k =0}^{n} {2n+1\choose 2k} 
% m_{2k} b_{n-k} \mu ^{2n-2k}\, .
\ee

The moments $m_{2k}$ are determined by integrating the density, which at the same time contain $m_{2k'}$.
In turn, this will give a linear system of equations for $m_{2k}$ that can be solved
in terms of $\mu $ and $\lambda $. Then we shall use the normalization condition to
compute $\mu $ in terms of $\lambda$. 

For example, if we wish to determine correlators up to $\lambda^8$ corrections, relative to the leading term, then we  need to truncate the $K$ series up to the term with coefficient $\zeta(9)$, which implies
computing up to $m_{8}$, i.e. solving a linear system for $m_2, m_4, m_6, m_8$ in terms of $\mu$ and $\lambda $. The linear system is obtained form the formula:
\be\label{equis}
m_{2r} =\sum_{k=0}^\infty C_k \int_{-\mu}^\mu dz\ z^{2k+2r} \sqrt{\mu^2-z^2} =
\pi \sum_{k=0}^\infty C_k \aaa_{k+r+1} \mu^{2k+2r+2}\, .
\ee

\subsubsection*{Next to leading order terms (NLO)}

In this paper we will compute the first non-trivial order in $\lambda $, that is, up to next-to-leading order (NLO). This is obtained by truncating the Taylor series
for $K$ at $n=1$. So we have:
\be
\label{arero}
A_0= - \frac{1}{\pi}  \zeta(3) \mu^{2}\, ;\qquad A_1=   - \frac{2}{\pi}  \zeta(3)  \, ;
\ee
\be
\label{berero}
B_{0} =\frac{1}{\pi} \zeta(3)   ( \mu^2 + 6 m_2 )\, ;\quad
B_1= \frac{2}{\pi} \zeta(3) \, .
\ee
As a warm-up, consider the undeformed theory.  We set $n_0=1$ (i.e. we set to zero all deformations
$g_{2n}$, $n>1$).
Then 
\be
q_0= \frac{1}{\pi}   g_{2} \ ,\ 
\ee
and all other $q_k=0$, $k>0$.
Thus

\bea
C_0 &=&   \frac{1}{\pi}   g_{2}  - \frac{1}{\pi}  \zeta(3) \mu^{2}+ \frac{1}{\pi} \zeta(3)   ( \mu^2 + 6 m_2 ) = \frac{8\pi}{\lambda } + \frac{6}{\pi} \zeta(3)  m_2 \, ;\\
C_1&=&  - \frac{2}{\pi}  \zeta(3)  +  \frac{2}{\pi} \zeta(3) =0\, .
\eea
Moreover, we have from (\ref{equis})
\be
m_2= \pi C_0 \aaa_{2} \mu^{4} =  (\frac{\pi^2}{\lambda } +\frac{3}{4} \zeta(3)  m_2 )\mu^{4} \, .
\ee
These results agree with those in \cite{Passerini:2011fe}.

Let us now restore the deformations. We consider  (\ref{equis}), with  $A_0,\ A_1$ and $B_0,\ B_1$ given by (\ref{arero}) and (\ref{berero}), while $q_k $  given by (\ref{qs}). Then
\be
\label{emedos}
m_2=\frac{3}{4} \zeta(3)  m_2  \mu^{4} + A\, ,
\ee
where
\bea
A\equiv \pi \sum_{k=0}^{n_0-1} q_k \aaa_{k+2} \mu^{2k+4} &=& \sum_{k=0}^\infty \sum_{m=k+1}^{n_0} m  b_{m-k-1} \aaa_{k+2}  g_{2m} \mu^{2m+2} 
\nonumber\\
 &=& \sum_{m=1}^{n_0} \eta_m  g_{2m} \mu^{2m+2}\, ;
\eea
with
\be
\eta_m \equiv \sum_{k=0}^{m-1} m  b_{m-k-1} \aaa_{k+2} = \frac{m^2\Gamma(m+\frac{1}{2})}{2\sqrt{\pi} (m+1)!}\, .
\ee

The various correlations functions can now be obtained from differentiating the free energy $F$
with respect to the coupling $g_{2n}$ a certain number times. As in the ${\cal N}=4$ case, 
we begin with 
\be
\label{cui}
\partial_{g_{2n}} F= \int_{-\mu}^\mu dz \ z^{2n} \rho(z) =m_{2n}\, .
\ee
This gives
\be
\label{coi}
\partial_{g_{2n}} F = \pi \sum_{k=0}^{n_0} C_k \aaa_{k+n+1}  \mu^{2k+2n+2} 
= \sum_{m=1}^{n_0} d_{m,n} g_{2m} \mu^{2m+2n} +  6\zeta(3)m_2 \aaa_{n+1}  \mu^{2n+2} \, ,
\ee
with $d_{m,n}$ given earlier in (\ref{ddmn}). Thus, 
\bea
\label{arfe}
\partial_{g_{2j}}\partial_{g_{2n}} F 
&=& \sum_{m=1}^{n_0}  (2m+2n) d_{m,n} g_{2m} \mu^{2m+2n-1} \frac{d\mu}{dg_{2j}}+ d_{j,n} \mu^{2j+2n} 
\nonumber\\
&+&  6\zeta(3)m_2 (2n+2) \aaa_{n+1}  \mu^{2n+1}  \frac{d\mu}{dg_{2j}}+ 6\zeta(3) \aaa_{n+1}  \mu^{2n+2} \frac{dm_2}{dg_{2j}}\, .
\eea
The derivatives $ \frac{d\mu}{dg_{2r}}$, $\frac{dm_2}{dg_{2r}}  $ are computed in the appendix \ref{details}, with the result
\be
\label{vmup}
\frac{d\mu}{dg_{2r}}  = -  \mu^{2r+1}\frac{4r b_r+3 \zeta(3) (4\eta_r- r b_r)\mu^4 }{8+6\zeta(3)\mu^4}\ .
\ee
\be
\label{vmp}
\frac{dm_2}{dg_{2r}} = \frac{ \eta_r\mu^{2r+2}+\mu \frac{d\mu}{dg_{2r}} }{1- \frac{3}{4}\zeta(3)\mu^4}= - \mu^{2r+2} \frac{r \aaa_{r+1}}{1+ \frac{3}{4}\zeta(3)\mu^4}
\, .
\ee
Substituting into (\ref{arfe}), we  find
\be
\label{totalF}
\partial_{g_{2m}}\partial_{g_{2n}} F = -\mu^ {2m+2n} \frac{\Gamma(m+\frac{1}{2})\Gamma(n+\frac{1}{2}) }{\pi (m+n) \Gamma(m) \Gamma(n) } Q_{m,n}\, ,
\ee
with
\be 
Q_{m,n}\equiv \frac{1}{1+ \frac{3}{4}\zeta(3)\mu^4}\left(1+ \frac{3}{4}\zeta(3)\mu^4\frac{ (m-1)(n-1)}{(m+1)(n+1)}\right)
\ee
On the other hand, from the normalization condition, one finds (see (\ref{vnor}))
\be
\label{mu}
1=\mu^2\left( \frac{4\pi^2}{\lambda} +  \frac{3}{4}\zeta(3)\mu^2\right)\ .
\ee
Solving for $\mu^2$, substituting into (\ref{totalF}) and expanding in powers of $\lambda $, we find
\be
\partial_{g_{2m}}\partial_{g_{2n}} F =\partial_{g_{2m}}\partial_{g_{2n}} F_0 + P^{(2)}_{m,n}
+ P^{(4)}_{m,n} +...\, ;
\ee
with
\be
\label{leading}
\partial_{g_{2m}}\partial_{g_{2n}} F_0 = -\left( \frac{\lambda}{4\pi^2} \right)^ {m+n} \frac{\Gamma(m+\frac{1}{2})\Gamma(n+\frac{1}{2}) }{\pi (m+n) \Gamma(m) \Gamma(n) }\, ,
\ee
\be
P_{m,n}^{(2)} =\frac{3}{4}\zeta(3) \left(\frac{\lambda}{4\pi^2}  \right)^ {m+n+2} \frac{(3 + m + n + m n) \Gamma(m+\frac{1}{2})\Gamma(n+\frac{1}{2}) }{\pi (m+1)(n+1) \Gamma(m) \Gamma(n) }\, ,
\ee
\be
P_{m,n}^{(4)} =-\frac{9}{32}\zeta(3)^2 \left(\frac{\lambda}{4\pi^2}  \right)^ {m+n+4} \frac{(3 + m + n) (5 + m + n + m n)\Gamma(m+\frac{1}{2})\Gamma(n+\frac{1}{2}) }{\pi (m+1)(n+1) \Gamma(m) \Gamma(n) }\, .
\ee
Note that $P_{m,n}^{(2)}$ is a correction of $O(\lambda^2)$ relative to the leading term and it corresponds to an extra loop (see \cite{Penati:1999ba} for the diagrammatic). Moreover, the leading term \eqref{leading} is just the same as that for the $\mathcal{N}=4$ case in eq.\eqref{eveneven2}. 
This is to be expected, since the leading term corresponds to the free theory; while, in turn, due to the non-renormalization theorem \cite{Freedman:1998tz,D'Hoker:1998tz,Penati:1999ba}, the full $\mathcal{N}=4$ result is just given by the free theory.
To order  $O(\lambda^3)$, there are extra contributions proportional to $\zeta(5)$, in addition to $P_{m,n}^{(4)}$, which we are not computing. However,  the above formulas encapsulate the complete dependence on  $\zeta(3)$ in the two-point functions to all order in the coupling (for terms involving powers of $\zeta(3)$ with no other factor $\zeta(2n+1)$, $n>1$).

It is interesting to note that, for very large $n,m$, $Q_{m,n}\rightarrow 1$. Nevertheless note that $\mu$ is determined from eq. \eqref{mu}, which differs from the $\mathcal{N}=4$ SYM case in the $\mathcal{O}(\zeta(3)\lambda)$ term. Thus, even in this limit, there will still be non-trivial NLO corrections.

\subsubsection*{Gram-Schmidt orthogonalization}

Having computed  the second derivatives of the free energy, we can now proceed to perform the Gram-Schmidt orthogonalization procedure. Following the same steps as above, the first few orthogonalized operators are (to next-to-leading order in $\lambda$)

\begin{eqnarray}
\label{OperatorsSQCD}
&&O_2^{S^4}\,; \nonumber \\
&&O_4^{S^4}-\frac{4\lambda}{(4\pi)^2}\Big(1-\frac{3\zeta(3)}{64\pi^4}\lambda^2+\mathcal{O}(\lambda^3)\Big)\,O_2^{S^4}\, ;\nonumber \\
&&O_6^{S^4}-\frac{6\lambda}{(4\pi)^2}\Big(1-\frac{3\zeta(3)}{64\pi^4}\lambda^2+\mathcal{O}(\lambda^3)\Big)O_4^{S^4}+\frac{9\lambda^2}{(4\pi)^4}\Big(1-\frac{3\zeta(3)}{32\pi^4}\lambda^2+\mathcal{O}(\lambda^3)\Big)O_2^{S^4}\, ; \nonumber \\
&&O_8^{S^4}-\frac{8\lambda}{(4\pi)^2}\Big(1-\frac{3\zeta(3)}{64\pi^4}\lambda^2+\mathcal{O}(\lambda^3)\Big)O_6^{S^4}+\frac{20\lambda^2}{(4\pi)^4}\Big(1-\frac{3\zeta(3)}{32\pi^4}\lambda^2+\mathcal{O}(\lambda^3)\Big)O_4^{S^4}\nonumber \\ && \nonumber \hspace{7cm}-\frac{16\lambda^3}{(4\pi)^6}\Big(1-\frac{9\zeta(3)}{64\pi^4}\lambda^2+\mathcal{O}(\lambda^3)\Big)O_2^{S^4}\, ; \nonumber \\
&&\cdots 
\end{eqnarray}
where $O_n^{S^4}$ are the VEV-less operators. As expected, these operators coincide, at leading order, with the operators in the $\mathcal{N}=4$ theory in eq.\eqref{OperatorsN=4}. Taking into account numerical factors as described above, the correlators for $O_n^{\mathbb{R}^4}={\rm Tr}\phi^n$ on $\mathbb{R}^4$ are given by 

\begin{eqnarray}
\label{SCQDCPO}
\langle O_2^{\mathbb{R}^4}(0)\overline{O}_{\overline{2}}^{\mathbb{R}^4}(x)\rangle_{\mathbb{R}^4}&=&\frac{1}{|x|^{4}}\frac{2\lambda^2}{(2\pi)^4}\Big(1-\frac{9\zeta(3)}{4 (2\pi)^4}\lambda^2+\mathcal{O}(\lambda^3)\Big)\, ; \nonumber \\
\langle O_4^{\mathbb{R}^4}(0)\overline{O}_{\overline{4}}^{\mathbb{R}^4}(x)\rangle_{\mathbb{R}^4}&=&\frac{1}{|x|^{8}} \frac{4\lambda^4}{(2\pi)^8}\Big(1-\frac{3\zeta(3)}{(2\pi)^4}\lambda^2+\mathcal{O}(\lambda^3)\Big)\,;\nonumber\\
\langle O_6^{\mathbb{R}^4}(0)\overline{O}_{\overline{6}}^{\mathbb{R}^6}(x)\rangle_{\mathbb{R}^4}&=&\frac{1}{|x|^{12}} \frac{6\lambda^6}{(2\pi)^{12}}\Big(1-\frac{9\zeta(3)}{2(2\pi)^4}\lambda^2+\mathcal{O}(\lambda^3)\Big)\,;\nonumber\\
\langle O_8^{\mathbb{R}^4}(0)\overline{O}_{\overline{8}}^{\mathbb{R}^6}(x)\rangle_{\mathbb{R}^4}&=&\frac{1}{|x|^{16}} \frac{8\lambda^8}{(2\pi)^{16}}\Big(1-\frac{6\zeta(3)}{(2\pi)^4}\lambda^2+\mathcal{O}(\lambda^3)\Big)\,;\nonumber\\
&\cdots&
\end{eqnarray}
Let us consider in detail the $\langle O_2^{\mathbb{R}^4}(0)\overline{O}_{\overline{2}}^{\mathbb{R}^4}(x)\rangle_{\mathbb{R}^4}$ correlator. Writting it in terms of $N$, it reads

\begin{equation}
\label{tures}
\langle O_2^{\mathbb{R}^4}(0)\overline{O}_{\overline{2}}^{\mathbb{R}^4}(x)\rangle_{\mathbb{R}^4}=\frac{1}{|x|^4}\Big(\frac{2N^2}{\pi^2 ({\rm Im}\tau_{YM})^2}-\frac{9N^4\zeta(3)}{2\pi^4({\rm Im}\tau_{YM})^4}+...\,\Big).
\end{equation}
Combining with the results for $SU(2),\ SU(3),\ SU(4)$ gauge groups discussed in \cite{Gerchkovitz:2016gxx} and in \cite{Baggio:2014ioa}, 
it is natural to conjecture that the finite $N$ version of  formula (\ref{tures}) is

\begin{equation}
\langle O_2^{\mathbb{R}^4}(0)\overline{O}_{\overline{2}}^{\mathbb{R}^4}(x)\rangle_{\mathbb{R}^4}=\frac{1}{|x|^4}\Big(\frac{2(N^2-1)}{\pi^2 ({\rm Im}\tau_{YM})^2}-\frac{9(N^2-1)(N^2+1)\zeta(3)}{2\pi^4({\rm Im}\tau_{YM})^4}+...\Big)\,,
\end{equation}
so that the ratio of the NLO to the leading order is $9(N^2+1)\frac{\zeta(3)}{4\pi^2\,({\rm Im}\tau_{YM})^2}$. 
For $N=2$, this formula perfectly agrees with eqs. (3.24) in \cite{Gerchkovitz:2016gxx} and eq. (5.35) in \cite{Baggio:2014ioa}; while for $N=3$ and $N=4$ it reproduces (3.43) and (3.54) in \cite{Gerchkovitz:2016gxx} respectively. 

It is interesting to note that including next-to-NLO in the correlators above will bring contributions of the form $\zeta(5)\lambda^3$. Yet, since $K$ only contains $\zeta(2n+1)$, interesting contributions of order $\lambda^4$ --in particular  $\zeta(3)^2\lambda^4$ terms-- can be read from the computation which we have performed. Explicitly showing the structure of the corrections we have

\begin{eqnarray}
\langle O_2^{\mathbb{R}^4}(0)\overline{O}_{\overline{2}}^{\mathbb{R}^4}(\infty)\rangle_{\mathbb{R}^4}&=&\frac{1}{|x|^4}\frac{2\lambda^2}{(2\pi)^4}\Big\{1-\frac{9\zeta(3)}{4 }\frac{\lambda^2}{(2\pi)^4}+a_2\,\zeta(5)\frac{\lambda^3}{(2\pi)^6} \nonumber\\ && \hspace{1cm}+\Big(\frac{45\zeta(3)^2}{8}+n_2\zeta(7)\Big)\frac{\lambda^4}{(2\pi)^8}+\mathcal{O}(\lambda^5)\Big\}\, ; \nonumber \\
\langle O_4^{\mathbb{R}^4}(0)\overline{O}_{\overline{4}}^{\mathbb{R}^4}(\infty)\rangle_{\mathbb{R}^4}&=&\frac{1}{|x|^8}\frac{4\lambda^4}{(2\pi)^8}\Big\{1-3\zeta(3)\frac{\lambda^2}{(2\pi)^4}+a_4\,\zeta(5)\frac{\lambda^3}{(2\pi)^6}\nonumber  \\ && \hspace{1cm}+\Big(\frac{63\zeta(3)^2}{8}+n_4\zeta(7)\Big)\frac{\lambda^4}{(2\pi)^8}+\mathcal{O}(\lambda^5)\Big\}\,;\nonumber\\
\langle O_6^{\mathbb{R}^4}(0)\overline{O}_{\overline{6}}^{\mathbb{R}^6}(\infty)\rangle_{\mathbb{R}^4}&=& \frac{1}{|x|^{12}}\frac{6\lambda^6}{(2\pi)^{12}}\Big\{1-\frac{9\zeta(3)}{2}\frac{\lambda^2}{(2\pi)^4}+a_6\,\zeta(5)\frac{\lambda^3}{(2\pi)^6} \nonumber \\ && \hspace{1cm}+\Big(\frac{243\zeta(3)^2}{16}+n_6\zeta(7)\Big)\frac{\lambda^4}{(2\pi)^8}+\mathcal{O}(\lambda^5)\Big\}\,;\nonumber\\
\langle O_8^{\mathbb{R}^4}(0)\overline{O}_{\overline{8}}^{\mathbb{R}^6}(\infty)\rangle_{\mathbb{R}^4}&=&\frac{1}{|x|^{16}}\frac{8\lambda^8}{(2\pi)^{16}}\Big\{1-6\zeta(3)\frac{\lambda^2}{(2\pi)^4}+a_8\,\zeta(5)\frac{\lambda^3}{(2\pi)^6} \nonumber \\ && \hspace{1cm}+\Big(\frac{198\zeta(3)^2}{8}+n_8\zeta(7)\Big)\frac{\lambda^4}{(2\pi)^8}+\mathcal{O}(\lambda^5)\Big\}\,;\nonumber\\
&\cdots&
\end{eqnarray}
where $a_k,\ n_k$ are (rational) numerical coefficients. These can be determined by
the same method used in this paper, by including more terms in the Taylor expansion of $K(x)$.

\subsection{Comments on strong coupling}\label{StrongCoupling}

The matrix model for undeformed ${\cal N}=2$ superconformal QCD  was investigated at strong ($\lambda\gg 1 $) coupling in 
\cite{Passerini:2011fe,Bourgine:2011ie,Russo:2012ay}.
In the infinite $\lambda$ coupling limit, one can exactly solve the saddle-point  equation (\ref{ferp}).
In this limit, the harmonic potential $8\pi x^2/\lambda$ vanishes and the eigenvalue distribution extends from $-\infty $ to $\infty $. Then, a Fourier analysis gives the normalized density
 \cite{Passerini:2011fe}
\be
\rho_\infty (x) =\frac{1}{2\cosh \frac{\pi x}{2}} \ .
\ee
This immediately allows us to obtain the VEV (in the $S^4$) of chiral operators in the infinite coupling limit:
\be
\langle {\rm Tr}\phi^{2n} \rangle =\int_{-\infty}^\infty dx \ \rho(x) x^{2n} = |E_{2n}|
\ee
where $E_{2n}$ is the Euler number, while $\langle {\rm Tr}\phi^{2n+1} \rangle =0$.

The calculation of two-point functions requires going beyond this leading order. In the present case,
this requires a more sophisticated analysis, similar to the one carried out in \cite{Passerini:2011fe,Bourgine:2011ie,Russo:2012ay} to determine subleading contributions in $1/\lambda $ in the free energy.
For finite $\lambda \gg 1 $, the eigenvalue distribution has a width $\mu $ of the form
\cite{Passerini:2011fe}
\be
\mu= \frac{2}{\pi}\ln \lambda +...\ .
\ee
In Fourier space, one has 
\be
\rho(x)=\int_{-\infty}^\infty \frac{d\omega}{2\pi} \ e^{-i\omega x} \rho(\omega )\ ,
\ee
where 
\be
 \rho(\omega )=\frac{1}{\cosh\omega }+\frac{2\sinh^2\frac{\omega}{2}}{\cosh{\omega}} \ \frac{8\pi^2\mu J_1(\mu w)}{\lambda \omega}+ ....
\ee
Then, a slight generalization of the calculation in appendix B of \cite{Russo:2012ay} gives
\be
\partial_{g_{2n}} F= \langle {\rm Tr}\phi^{2n} \rangle = \int_{-\mu}^\mu dx \rho(x)x^{2n} =- \int_{-\infty}^\infty \frac{d\omega}{\pi i} \frac{\rho^{(2n)}(\omega )}{\omega -i0}\ . 
\ee
Application of the Sokhotski-Plemelj formula then determines $\langle {\rm Tr}\phi^{2n} \rangle $
to be proportional to the  $2n$-th derivative of $\rho(\omega )$ at $\omega=0$.
This expresses  $\partial_{g_{2n}} F$ in terms of $\mu $, hence in terms of $\lambda $.
Using this result, one can compute two-point functions $\partial_{g_2}\partial_{g_{2n}} F$.
The computation of  $\partial_{g_{2m}}\partial_{g_{2n}} F$ is more difficult and requires
generalizing the analysis of  \cite{Passerini:2011fe} to the case of a potential $V=g_{2m}x^{2m}$.
It would be extremely interesting to determine the general two-point functions in the
strong coupling limit.

\section{Conclusions}\label{conclusions}

Using large $N$ techniques we have computed correlation functions for CPO's in $\mathcal{N}=4$ SYM and superconformal QCD. For large $N$, the CPO's of interest are single-trace operators of the form ${\rm Tr}\phi^n$. When the theory is on $S^4$, due to the conformal anomaly, operators corresponding to different $n$'s mix. As we have argued, such mixing occurs only among operators ${\rm Tr}\phi^n$, ${\rm Tr}\phi^m$, having $n$ and $m$ of the same parity.
This reflects the fact that the mixing is due to the conformal anomaly, which, through the Ricci scalar $R$, allows mixings of operators whose dimensions differ by 2 (see (\ref{mixings})).

In the case of $\mathcal{N}=4$ SYM the final outcome of the computation is encoded in eq. \eqref{N=4CPO}. This formula nicely agrees with the results in the literature computed in perturbation theory. Indeed, the result is correct at leading order in the free field theory case, as it has been argued that, due to a non-renormalization theorem, higher corrections cancel. Our computation explicitly confirms this. It is also worth noting that, instead of using the deformed
matrix model, another, more direct approach to the computation of correlation functions
of CPO's is by directly computing the matrix integrals of the Gaussian matrix model with insertions of ${\rm Tr} a^{n_i}$ by using orthogonal polynomials. Like in the case of the Wilson loop \cite{Drukker:2000rr}, this approach may also permit to  find closed expressions for any finite $N$.
We plan to report on this approach in a future publication.

In the case of superconformal QCD we have considered the large $N$, small $\lambda$ regime, which is akin to the planar expansion. The leading order is identical to the $\mathcal{N}=4$ SYM theory. Due to the non-renormalization theorem, this is in turn simply given by the free theory, which should be indeed giving the leading term in superconformal QCD as well. We computed the NLO correction, which admits a
 simple extension into a finite $N$ formula which exactly matches small $N$ results in the literature. We have computed up $O(\lambda^2)$ with respect to the leading order. It would be definitely interesting to
compute higher order in $\lambda $ and understand the systematic of the expansion. Note however that, thanks to the transcendentality properties of the perturbative expansion, we also computed the next-to-next-to-NLO proportional to $\zeta(3)^2\lambda^4$. It would also be extremely interesting to compute two-point functions in the strong $\lambda\gg 1$ limit, following the method outlined
 in section \ref{StrongCoupling}.

\section*{Acknowledgements}

We would like to thank G.Bonelli for very useful discussions.
D.R-G si partly supported by the Ramon y Cajal grant RYC-2011-07593, the asturian grant FC-15-GRUPIN14-108 as well as the EU CIG grant UE-14-GT5LD2013-618459. D.R-G. would like to thank the U. Barcelona for warm hospitality during the initial stages of this project. 
J.G.R. acknowledges financial support from projects  FPA2013-46570,  
 2014-SGR-104 and  MDM-2014-0369 of ICCUB (Unidad de Excelencia `Mar\'ia de Maeztu').

\begin{appendix}

\section{Decoupling of multi-trace operators}\label{DecouplingOfMultitrace}

Since in the present $S^4$ case there is a non-trivial operator mixing, it is useful to understand the decoupling of multitrace operators in more detail. In the large $N$ limit, correlation functions involving multitrace operators are known to be suppressed by extra powers of $1/N$.
In short, this can be argued by noting that the large $N$ free energy is of the form
$\mathcal{F}(\tau_k,\overline{\tau}_k)= -N^2 F|_{\rm saddle}(\mathfrak{g}_k,\bar {\mathfrak{g}}_k)$, which exhibits explicitly the $N$ dependence.
Correlators  are obtained by differentiating with respect to $\partial_{\tau_n} $, but 
$\partial_{\tau_n} \sim N^{-1} \partial_{\mathfrak{g}_n} $, see (\ref{calg}).
Since each $\partial_{\tau_n}$ inserts a single-trace operator, it is clear that multitrace operators require
more $\partial_{\tau_n}$ derivatives and therefore additional $1/N$ factors
(for example, one may compare correlators involving ${\rm Tr}\phi^6$ with 
correlators involving ${\rm Tr}\phi^2{\rm Tr}\phi^4$ - the latter will carry an extra $1/N$ factor). Nevertheless, let us be more specific in our argumentation (as a by-product, we will make more explicit the Gram-Schmidt orthogonalization preocedure).  Let us suppose that, up to order $\Delta_0$, we have constructed all the correlators of the VEV-less operators $O^{S^4}_{n_1,\cdots,n_N}$, which in particular implies $\sum m\,n_m\leq\Delta_0$. Let us suppose that, with these, we have constructed the orthogonal basis of operators $O^{\mathbb{R}^4}_i$ --$i$ stands for an arbitrary chosen ordering of such operators--, covering all dimensions up to $\Delta_0$. Assume now that for these it is true that multi-trace operators decouple in the large $N$ limit; that is, the set of operators $O^{S^4}$ (or equivalently the $O^{\mathbb{R}^4}$) contains only single-trace operators. We will prove by induction that this holds at any order. To that matter, suppose now that we want to consider operators up to dimension $\Delta=\Delta_0+1$. Now the relevant operators to include in the orthogonalization are those satisfying $O^{S^4}_{n_2,\cdots,n_N}$with $\sum m\,n_m\leq\Delta_0+1$. Let us split them into

\begin{equation}
A=\{O^{S^4}_{n_1,\cdots,n_N}\,/\,\sum m\,n_m\leq\Delta_0\}\, ,\qquad B=\{O^{S^4}_{n_1,\cdots,n_N}\,/\,\sum m\,n_m=\Delta_0+1\}\, .
\end{equation}
The operators in $A$ are nothing but the set of all operators up to dimension $\Delta_0$ for which we are assuming that multi-traces decouple in the large $N$ limit. Therefore, $A$ contains only single-traces and moreover we can change basis into the orthogonalized $\{O^{\mathbb{R}^4}_n,\,n=1\cdots \Delta_0\}$ which satisfy $G_{n\overline{m}}\sim N^0\,\delta_{n\overline{m}}$.

As for the operators in $B$, let us choose some ordering for them so that $B=\{B_1,\cdots, B_M\}$. All these operators are of dimension $\Delta_0+1$, but only one of them will be a single trace operator. With no loss of generality, let us assume it to be $B_1$. The Gram-Schmidt orthognalization procedure amounts to write

\begin{equation}
\label{ODelta+1}
O^{\mathbb{R}^4}_{\Delta_0+1}=B_1-\sum_{n=1}^{\Delta_0}\alpha_nO^{\mathbb{R}^4}_i\, ,
\end{equation}
and fix the $\alpha_n$ coefficeints by demanding $\langle O^{\mathbb{R}^4}_{\Delta_0+1}\overline{O}^{\mathbb{R}^4}_i\rangle=0$ for all $i=1\cdots M$. Due to the orthogonality properties of the $O^{\mathbb{R}^4}_i$ we have

\begin{equation}
\label{coef}
\alpha_n=\frac{\langle B_1\overline{O}^{\mathbb{R}^4}_n\rangle}{\langle O^{\mathbb{R}^4}_n\overline{O}^{\mathbb{R}^4}_n\rangle}\, ,\qquad n=1,\cdots, \Delta_0\, .
\end{equation}
This way we find a set of $\Delta_0+1$ operators, and we can further proceed with $B_2$ and construct its associated orthogonal operator. Now

\begin{equation}
O^{\mathbb{R}^4}_{\Delta_0+2}=B_2-\sum_{n=1}^{\Delta_0+1}\alpha_nO^{\mathbb{R}^4}_n\qquad \alpha_n=\frac{\langle B_2\overline{O}^{\mathbb{R}^4}_n\rangle}{\langle O^{\mathbb{R}^4}_n\overline{O}^{\mathbb{R}^4}_n\rangle}\, ,\qquad n=1,\cdots, \Delta_0+1\, .
\end{equation}
Continuing in the same way we can find the set of orthogonal operators up to dimension $\Delta=\Delta_0+1$. Recall that, since by assumption the single-trace operator is $B_1$, only $O^{\mathbb{R}^4}_{\Delta_0+1}$ involves only single-traces, while the other  $O^{\mathbb{R}^4}_{\Delta_0+2}$ \textit{etc}. do involve multi-traces.

Let us now examine in closer detail the mixing coefficients in $O^{\mathbb{R}^4}_{\Delta_0+1}$ as opposed to those in $O^{\mathbb{R}^4}_{\Delta_0+2}$ \textit{etc}. (let us choose $O^{\mathbb{R}^4}_{\Delta_0+2}$ for concreteness).  Since only $B_1$ is single-trace, and by assumption the $O^{\mathbb{R}^4}_n$ for $n\leq \Delta_0$ are single-traces, schematically we have $\langle B_1\overline{O}^{\mathbb{R}^4}_n\rangle \sim \langle {\rm Tr}\overline{\rm Tr}\rangle$ while $\langle B_2\overline{O}^{\mathbb{R}^4}_n\rangle\sim \langle {\rm Tr}^n \overline{\rm Tr}\rangle$ and $\langle O^{\mathbb{R}^4}_n\overline{O}^{\mathbb{R}^4}_n\rangle\sim \langle {\rm Tr}^2\rangle$. 

The correlation functions of the $O^{S^4}$ operators are constructed by taking appropriate derivatives of the sphere partition function of the deformed theory. Recall now that in the computation of $\mathcal{F}$ we introduce a set of $\mathfrak{g}_n$'s appropriately rescaled (see main text) so that the $N$ dependence is extracted as $\mathcal{Z}=e^{-\mathcal{F}}$, with $\mathcal{F}=N^2F(\{\mathfrak{g}_n\})$. As mentioned before, from the relation between the $\mathfrak{g}_n$'s and the $\tau_n$ we have that, up to constant terms (including the 't Hooft coupling, fixed in the large $N$ limit)

\begin{equation}
\partial_{\tau_n}\sim\frac{1}{N} \partial_{g_n}\, .
\end{equation}
From these scalings we see that derivatives of the free energy scale with $N$ as (note that for these matters $\partial_{\tau}\sim\partial_{\overline{\tau}}$)

\begin{equation}
\partial_{\tau}^n\mathcal{F}\sim N^{2-n}\, .
\end{equation}
Using this scaling, we find that the scaling of $\langle {\rm Tr}^n\rangle$ is\footnote{One can convince of this very easily by explicitly looking at the first few examples
\begin{equation}
\langle {\rm Tr}\rangle=-\partial_{\tau}\mathcal{F}\, ,\qquad \langle {\rm Tr}^2\rangle=- \partial^2_{\tau}\mathcal{F}+(\partial_{\tau}\mathcal{F})^2\, ,\qquad \langle {\rm Tr}^3\rangle\sim \partial^3_{\tau}\mathcal{F}+\partial_{\tau}\mathcal{F} \partial_{\tau}^2F+(\partial_{\tau}\mathcal{F})^3\qquad \cdots
\end{equation}
Note that the leading term reflects to the familiar large $N$ factorization $\langle {\rm Tr}^n\rangle \sim \langle {\rm Tr}\rangle^n$.}

\begin{equation}
\langle {\rm Tr}^n\rangle\sim  N^n+\cdots+N^{2-n}\, .
\end{equation}
Therefore

\begin{equation}
\frac{\langle B_1\overline{O}_i\rangle}{\langle O_i\overline{O}_i\rangle} \sim \frac{\langle {\rm Tr}\overline{\rm Tr}\rangle}{\langle{\rm Tr}\overline{\rm Tr}\rangle}\sim 1\, ,\qquad \frac{\langle B_2\overline{O}_i\rangle}{\langle O_i\overline{O}_i\rangle}\sim \frac{\langle {\rm Tr}^n\overline{\rm Tr}\rangle}{\langle {\rm Tr}\overline{\rm Tr}\rangle}\sim   \frac{N^{n+1}+\cdots+N^{1-n}}{N^2+1}\sim N^{n-1}\, .
\end{equation}
Thus, schematically

\begin{equation}
O^{\mathbb{R}^4}_{\Delta_0+1}\sim B_1+\sum_{n}^{\Delta_0} \hat{\alpha}_n\,N^0\,O^{\mathbb{R}^4}_i\, ,\qquad O^{\mathbb{R}^4}_{\Delta_0+2}\sim B_2+\sum_{n}^{\Delta_0+1} \hat{\alpha}_n\,N^{M_2}O^{\mathbb{R}^4}_i\, ;
\end{equation}
where the $\hat{\alpha}_n$ are coefficients with no $N$ dependence, and $M_2\geq1$ (its precise value will be given by the number of traces in $B_2$ minus one). This motivates to eliminate this extra $N$ suppression in mixings by redefining $O^{\mathbb{R}^4}_{\Delta_0+2}\rightarrow N^{-M_2}O^{\mathbb{R}^4}_{\Delta_0+2}$. In this manner we recover the familiar extra $\frac{1}{N}$ suppression in multitrace operators. At the same time, it becomes evident that the $O^{\mathbb{R}^4}_{\Delta_0+2}$ correlators come with an extra factor of $N^{-2M_2}$. This explicitly shows the decoupling of multitraces at dimension $\Delta_0+1$. Since one can explicitly check that multi-traces decouple at the lowest $\Delta_0$, this serves as an inductive proof explicitly showing that indeed multi-trace operators can be consistently neglected in the large $N$ limit.

With the formulae above, we can also easily give a closed form for the $O^{\mathbb{R}^4}_{\Delta{0+1}}$ correlator. Using eqs. \eqref{ODelta+1} and \eqref{coef} we have

\begin{equation}
\label{correlatorSU(N)}
\langle O^{\mathbb{R}^4}_{\Delta{0+1}}\overline{O}^{\mathbb{R}^4}_{\Delta{0+1}}\rangle=\langle B_1\overline{B}_1\rangle-\sum_n^{\Delta_0}\frac{\langle B_1\overline{O}_n^{\mathbb{R}^4}\rangle \langle O_n^{\mathbb{R}^4}\overline{B}_1\rangle}{\langle O_n^{\mathbb{R}^4}\overline{O}_n^{\mathbb{R}^4}\rangle}\,.
\end{equation}

\section{Useful formulas}\label{details}

We collect some relevant details for the computations in the main text, including normalization conditions and cut endpoints.

\subsection{$\mathcal{N}=4$ SYM}

We compile relevant results for the computation of correlators in the $\mathcal{N}=4$ SYM theory.

\subsubsection{Derivatives of endpoints of eigenvalue distribution: case of even deformation}

 The derivatives $d\mu/dg_{2k}$ can be computed from the normalization condition.
For the sake of clarity in the formulas, it is convenient to introduce another coefficient $p_k$ defined by
\be
p_k = k b_k =\frac{\Gamma(k+\frac{1}{2})}{\sqrt{\pi} (k-1)!}\ .
\ee
Differentiating the normalization condition (\ref{normapar}) with respect to $g_{2k}$, we obtain
\be
p_{k}\mu^{2k} =-\frac{d\mu}{dg_{2k}}\sum_{n=1}^{n_0} 2n p_{n} g_{2n} \mu^{2n-1}\ .
\ee
Thus
\be
\frac{d\mu}{dg_{2k}}= -\frac{p_{k}\mu^{2k}}{\sum_{n=1}^{n_0} 2n  p_{n} g_{2n} \mu^{2n-1}}\ .
\ee
It is useful to compute also  the second and third derivative of $\mu$. Differentiating again, we find
\bea
\frac{d^2\mu}{dg_{2k} dg_{2j}}
&=&\frac{2(k+j) p_{k}p_j \mu^{2k+2j-1}}{(\sum_{n=1}^{n_0} 2n  p_{n} g_{2n} \mu^{2n-1})^2}\ 
\nonumber\\
&-&\frac{p_j p_{k}\mu^{2k+2j}}{(\sum_{n=1}^{n_0} 2n  p_{n} g_{2n} \mu^{2n-1})^3}(\sum_{n=1}^{n_0} 2n (2n-1) p_{n} g_{2n} \mu^{2n-2})\ .
\eea
Now we evaluate the first and second derivatives at $g_{2n}=0$, with $n=2,...,n_0$.
We get
\be
\frac{d\mu}{dg_{2k}}= -\frac{p_{k}\mu^{2k-1}}{ 2  p_{1} g_{2}}= -\frac{p_{k}\mu^{2k-1}}{  g_{2}}\ ,
\ee
\be
\frac{d^2\mu}{dg_{2k} dg_{2j}}
=\frac{(2k+2j-1) p_{k}p_j \mu^{2k+2j-3}}{ 4  p_{1}^2 g_{2}^2 }
=\frac{(2k+2j-1) p_{k}p_j \mu^{2k+2j-3}}{  g_{2}^2 }\ ,
\ee
where we used $p_1=1/2$.
At the end, we evaluate $\mu $ at $g_{2n}=0$, $n=2,...$, which just gives
\be
\mu ^2= \frac{2}{g_2}=\frac{\lambda}{4\pi^2} \ .
\ee
Similarly, we find
\be
\frac{d^3\mu}{dg_{2\ell} dg_{2k} dg_{2j}}
= -\frac{1}{g_2^3} \big( 4(k+j+\ell)(k+j+\ell -2)+3\big) p_{\ell} p_{k}p_j\ \mu^{2k+2j+2\ell -5}\ .
\ee

\subsubsection{Derivatives of endpoints of eigenvalue distribution: general case with even and odd deformations}

The endpoints of the cut ($-\nu,\mu)$  are determined from the condition that the resolvent
behaves as  $\omega(x)\sim 1/x$ at large $x$. This leads to the condition (see \textit{e.g.} \cite{Marino:2011nm})
\be
I_1\equiv  \sum_{n=2}^{N}n  g_n \int dx \frac{x^{n-1}}{\sqrt{(\mu-x)(x +\nu)}} =0\ .
\ee
For an even potential, this  condition just  implies $\mu-\nu =0$. The integral can be computed by choosing a contour that surrounds the cut $(-\nu,\mu)$ and computing the residue at infinity, by changing integration variable $x= 1/z$.
Let us compute the generic integral 
\be
J_n\equiv \int dx \frac{x^{n}}{\sqrt{(\mu-x)(x +\nu)}} =-i \int \frac{dz}{z} \frac{1}{z^n \sqrt{(1-\mu z)(1 +\nu z)}}\ ,
\ee
where $z=1/x$. Expanding around $z=0$,
\be
J_n= -i \sum_{k=0}^\infty \sum_{\ell=0}^\infty b_k b_\ell \mu^k (-\nu)^\ell \int \frac{dz}{z}  z^{k+\ell-n}  \ .
\ee
By residues,
\be
J_n= \pi \sum_{k=0}^n   b_k b_{n-k} \mu^k (-\nu)^{n-k}\ .
\ee
Thus we have the condition
\be
\label{ire}
\sum_{n=1}^{N} n g_n J_{n-1} =0\ ,
\ee
i.e.
\be
\label{iqu}
\sum_{n=1}^{N} n g_n \sum_{k=0}^{n-1}   b_k b_{n-1-k} \mu^k (-\nu)^{n-1-k} =0\ .
\ee
Differentiating with respect to $g_\ell $, we get
\bea
0 &=& \ell  \sum_{k=0}^{\ell-1}   b_k b_{\ell-1-k} \mu^k (-\nu)^{\ell-1-k}
+ \frac{d\mu}{dg_\ell}
\sum_{n=1}^{N} n g_n \sum_{k=1}^{n-1}  k  b_k b_{n-1-k} \mu^{k-1} (-\nu)^{n-1-k}
\nonumber\\
&-& \frac{d\nu}{dg_\ell}
\sum_{n=1}^{N} n g_n \sum_{k=0}^{n-2}  (n-1-k)  b_k b_{n-1-k} \mu^{k} (-\nu)^{n-2-k}\ .
\eea
This has to be evaluated at $g_n=0$ for $n\ne 2$ and $\mu ^2= \nu^2=\frac{2}{g_2}$.
Hence
\be
0 = -\ell (-1)^{\ell } \mu^{\ell -1}\sum_{k=0}^{\ell-1}   b_k b_{\ell-1-k} (-1 )^{k}
+ g_2 \big(\frac{d\mu}{dg_\ell}
      -\frac{d\nu}{dg_\ell}\big) \ .
\ee
Thus
\be
 g_2 \big(\frac{d\mu}{dg_\ell}
      -\frac{d\nu}{dg_\ell}\big) =\ell (-1)^{\ell } \mu^{\ell -1}\gamma_{\ell-1}=
-\ell \mu^{\ell -1}\gamma_{\ell-1}\ ,
\ee
where $\gamma_{m}$ was computed earlier  in (\ref{gamita}). In the last equality we used that
$\gamma_{\ell-1}$ is $\neq 0$ only for odd $\ell $.

We have an extra condition coming from normalization. 
The normalization condition is
\be
1= -\pi \sum_{k=0}^{N-2} c_k\sum_{r=0}^{k+2}
\aaa_r \aaa_{k+2-r} \mu^r (-\nu)^{k+2-r}\ .
\ee
Differentiating with respect to $g_s$,  using (\ref{cuarenta}) (and  setting 
$\mu=\nu$ and $g_{n\ne2}=0$ after differentiation) we finally  obtain 
\be
 g_2  ( \partial_{g_s} \mu+ \partial_{g_s} \nu)=\mu^{s-1} h_s  \ ,
\label{cincuentas}
\ee
with
\be
h_s\equiv s\sum_{k=0}^{s-2} \alpha_{k+2}\gamma_{s-k-2}=-(1+(-1)^s) p_{\frac{s}{2}}\ .
\ee

\subsection{Superconformal ${\cal N}=2$ QCD}

We now compile several useful results for the computation of the correlators in the superconformal QCD case. 

The normalization condition can be read from (\ref{cui}), (\ref{coi}), setting $m_0=1$.
Using, in addition, $d_{m,0}=mb_m=p_m$, we have 
\be
1 =  \sum_{n=1}^{n_0}p_n g_{2n} \mu^{2n}+3 \zeta(3)  m_2   \mu^{2}\ .
\ee
Using
\be
\label{mui}
m_2=\frac{A}{1-\frac{3}{4} \zeta(3)\mu^4}\  ,
\ee
we finally find the condition
\be
0=(1-\frac{3}{4} \zeta(3)\mu^4) \sum_{n=1}^{n_0} p_n g_{2n} \mu^{2n}+3 \zeta(3)\sum_{n=1}^{n_0} \eta_n  g_{2n} \mu^{2n+4} -(1-\frac{3}{4} \zeta(3)\mu^4) \ .
\ee
Differentiating with respect to $g_{2r}$, we get
\bea
0 &=&(1-\frac{3}{4} \zeta(3)\mu^4)  p_r  \mu^{2r}+3 \zeta(3) \eta_r   \mu^{2r+4}+\sum_{n=1}^{n_0} 2n p_n g_{2n} \mu^{2n-1}\frac{d\mu}{dg_{2r}} 
\nonumber\\
&+ & 3\zeta(3)  \sum_{n=1}^{n_0} (\eta_n -\frac{1}{4} p_n) (2n+4)  g_{2n} \mu^{2n+3}\frac{d\mu}{dg_{2r}} 
+  3 \zeta(3) \mu^3 \frac{d\mu}{dg_{2r}} \ .
\eea
Setting all $g_{2n>2}=0$, and using $\eta_1= \frac{1}{4} p_1$, we obtain
\be
\frac{d\mu}{dg_{2r}}  = - \mu^{2r+1}\ \frac{4p_r+3 \zeta(3) (4\eta_r- p_r)\mu^4  }{4(g_2\mu^2+3\zeta(3)\mu^4)}\ .
\ee
We also need the derivative of $m_2$ with respect to $g_{2j}$.
Using (\ref{mui}), we find
\be
\frac{dm_2}{dg_{2j}} = \frac{2 \eta_j\mu^{2j+2}+(g_2 + 6\zeta(3) m_2 )\mu^3 \frac{d\mu}{dg_{2j}} }{2- \frac{3}{2}\zeta(3)\mu^4}\, .
\ee
These expressions can be further simplified by using the normalization condition, which relates $\mu^2$ to $g_2$. Upon setting $g_{2n>2}=0$, the normalization condition becomes
\be
\label{vnor}
1=\frac{1}{2} g_2\mu^2+3\zeta(3) m_2\mu^2\ ,
\ee
and $m_2=\mu^2/4$. Using these relations, we find the formulas (\ref{vmup}), (\ref{vmp}) given in section 4.

\section{Higher derivatives of the free energy in the $\mathcal{N}=4$ SYM theory}\label{HigherDerivatives}

Higher derivatives of the free energy generate correlators with higher number of
insertions. Since these correlators do not depend on the point, at the end these
correlators reduce to two-point functions of multitrace operators.
In this appendix we include formulas which might be relevant for higher point functions of even operators.

Three-point correlators can be obtained from the third derivative of the free energy:
\bea
\partial_{g_{2\ell}} \partial_{g_{2j}}\partial_{g_{2n}} F  &=& \sum_{m=1}^{n_0}   d_{m,n} g_{2m} (2m+2n) \mu^{2m+2n-2} \left( (2m+2n-1) \frac{d\mu}{dg_{2j}} \frac{d\mu}{dg_{2\ell }}   +\mu  
 \frac{d^2\mu}{dg_{2\ell} dg_{2j}}\right)
\nonumber\\
&+& d_{j,n} (2j+2n) \mu^{2j+2n-1} \frac{d\mu}{dg_{2\ell}}+d_{\ell ,n} (2\ell+2n) \mu^{2\ell+2n-1} \frac{d\mu}{dg_{2j}}\ .
\eea

At $g_{2m}=0,\ m> 1$, this simplifies to
\bea
\partial_{g_{2\ell}} \partial_{g_{2j}}\partial_{g_{2n}}   F &=& \frac{\mu^{2n+2j+2\ell-2}}{g_2}\bigg(  d_{1,n} p_j p_\ell  (2n+2) \left( 2\ell+2j +2n\right)
\nonumber\\
&-& d_{j,n} (2j+2n) p_\ell -d_{\ell ,n} (2\ell+2n) p_j \bigg)\ .
\eea

The  coefficients $d_{j,k}, \ p_j$ involve a combination of $\Gamma $ functions, and the three-point correlator finally becomes  the remarkably simple formula:
\be
\label{3point}
\partial_{g_{2\ell}} \partial_{g_{2j}}\partial_{g_{2n}}  F  = \left( \frac{\lambda}{4\pi ^2}  \right)^ {n+j+\ell }  \frac{\Gamma(\ell+\frac{1}{2})\Gamma(j+\frac{1}{2}) \Gamma(n+\frac{1}{2}) }{\pi^{\frac{3}{2}} \Gamma(\ell) \Gamma(j) \Gamma(n) }\ .
\ee

Four-point correlators are obtained from the fourth-derivative of the free energy
\bea
 \partial_{g_{2s}}\partial_{g_{2\ell}} \partial_{g_{2j}}\partial_{g_{2n}}  F  &=& 
\sum_{m=1}^{n_0}   d_{m,n} g_{2m} (2m+2n)  
\mu^{2m+2n-3}
\nonumber\\
&\times &\bigg[(2m+2n-1) \left((2m+2n-2)\frac{d\mu}{dg_{2s}}  \frac{d\mu}{dg_{2j }} \frac{d\mu}{dg_{2\ell }}   +\mu  \frac{d\mu}{dg_{2s}}
 \frac{d^2\mu}{dg_{2\ell } dg_{2j }}\right) 
\nonumber\\
&+ & \left( (2m+2n-1)\mu \Big(\frac{d^2\mu}{dg_{2j } dg_{2s}} \frac{d\mu}{dt_{\ell}}  +
\frac{d^2\mu}{dg_{2\ell } dg_{2s}} \frac{d\mu}{dg_{2j }} \Big)
 +\mu^2   \frac{d^3\mu}{dg_{2s} dg_{2\ell} dg_{2j }} \right)\bigg]
\nonumber\\
&+& d_{j,n} (2j+2n)\mu^{2j+2n-2}\left( (2j+2n-1) \frac{d\mu}{dg_{2\ell }}\frac{d\mu}{dg_{2s}}+\mu \frac{d^2\mu}{dg_{2\ell } dg_{2s}}\right)
\nonumber\\
&+& d_{\ell,n} (2\ell +2n)\mu^{2\ell +2n-2}\left( (2\ell+2n-1) \frac{d\mu}{dg_{2j }}\frac{d\mu}{dg_{2s}}+\mu \frac{d^2\mu}{dg_{2j } dg_{2s}}\right)
\nonumber\\
&+& d_{s,n} (2s+2n) \mu^{2s+2n-2} \left( (2s+2n-1) \frac{d\mu}{dg_{2j }} \frac{d\mu}{dg_{2\ell }}   +\mu  
 \frac{d^2\mu}{dg_{2\ell } dg_{2j }}\right)
\nonumber
\eea
Setting $g_{2m}=0,\ m>1$, we obtain
\bea
 \partial_{g_{2s}}\partial_{g_{2\ell}} \partial_{g_{2j}}\partial_{g_{2n}} F  &=& \frac{\mu^{2n+2s+2j+2\ell-4}}{g_2^2}
\bigg[ - d_{1,n}p_j p_s p_{\ell } (2n+2) 
\nonumber\\
&\times &\bigg((2n+1) \big(2n + 4\ell+4j+4s-3 \big)  +4(s+\ell+j)(s+\ell+j-2)+3\bigg)
\nonumber\\
&+& 4( j+n +\ell+s-1) \Big(d_{j,n} (j+n) p_{\ell}p_s 
+ d_{\ell,n} (\ell +n)p_{j}p_s 
+ d_{s,n} (s+n)p_{\ell }p_j  \Big)\bigg]
\nonumber
\eea

Using the expressions for the coefficients $d_{m,n}$, $p_k$, we find the remarkably simple formula

\be
 \partial_{g_{2s}}\partial_{g_{2\ell}} \partial_{g_{2j}}\partial_{g_{2n}}F =- \left( \frac{\lambda}{4\pi^2}  \right)^{n+s+j+k}
\frac{\Gamma \left(j+\frac{1}{2}\right) \Gamma \left(k+\frac{1}{2}\right) \Gamma
   \left(n+\frac{1}{2}\right) \Gamma \left(s+\frac{1}{2}\right) (j+k+n+s-1)}{\pi ^2
   \Gamma (j) \Gamma (k) \Gamma (n) \Gamma (s)}
\ee

\end{appendix}

\end{document}